\documentclass[useAMS,usenatbib,times,letter,amssymb]{mn2e}
\usepackage{graphics,latexsym,epsfig,graphicx,amssymb,lscape,color}     

\renewcommand{\d}[0]{{\rm d}}
\newcommand{\ave}[1]{\langle #1 \rangle}
\newcommand{\Ave}[1]{\Big\langle #1 \Big\rangle}
\newcommand{\Ref}[1]{\ref{#1}}

\renewcommand{\vec}[1]{{\bmath{#1}}}
\newcommand{\mat}[1]{\mathbfss{#1}}
\newcommand{\pprime}[0]{{\prime\prime}}
\newcommand{\msol}[0]{{\rm M}_\odot}

\title[3D analysis of HST/STAGES] {Spatial matter density
  mapping of the STAGES Abell A901/2 supercluster field with 3D
  lensing} \author[Simon et al.]  {P. Simon$^{1,2}$, C. Heymans$^2$,
  T. Schrabback$^3$, A.N. Taylor$^2$, M.E. Gray$^4$,\newauthor L. van
  Waerbeke$^{5}$, C. Wolf$^{6}$, D. Bacon$^7$,
M.  Barden$^{8}$,
  A. B\"ohm$^{8}$, 
B. H\"au\ss ler$^4$, \newauthor K.  Jahnke$^{9}$,
  S. Jogee$^{10}$, E. van Kampen$^{11}$, 
K.  Meisenheimer$^{9}$, C.Y.  Peng$^{12}$ \\
  $^1$Argelander-Institut f\"ur Astronomie, Universit\"at Bonn, Auf
  dem H\"ugel 71,
  53121 Bonn, Germany.\\
  $^2$The Scottish Universities Physics Alliance (SUPA), Institute for
  Astronomy,
  University of Edinburgh, Edinburgh EH9 3HJ, UK.\\
  $^3$Leiden Observatory, Leiden University, Niels Bohrweg 2, NL-2333 CA Leiden, The Netherlands.\\
  $^4$School of Physics and Astronomy, The University of Nottingham,
  University Park, Nottingham NG7 2RD, UK.\\
  $^5$Department of Physics and Astronomy, University of British
  Columbia, 6224 Agricultural
  Road, Vancouver, V6T 1Z1, Canada.\\
  $^{6}$Department of Physics, Denys Wilkinson
  Building, University of Oxford, Keble Road, Oxford, OX1 3RH, UK.\\
  $^7$Institute of Cosmology and Gravitation, University of
  Portsmouth, Dennis Sciama Building, Burnaby Road, Portsmouth, PO1 3FX, UK. \\
  $^{8}$Institute for Astro- and Particle Physics, University of
  Innsbruck, Technikerstr. 25/8, A-6020 Innsbruck, Austria. \\
  $^{9}$Max-Planck-Institut f\"{u}r Astronomie, K\"{o}nigstuhl 17,
  D-69117, Heidelberg,
  Germany.\\
  $^{10}$University of Texas, McDonald
  Observatory, Fort Davis, TX 79734, USA.\\
  $^{11}$European Southern Observatory, Karl-Schwarzschild-Str 2,
  D-85748
  Garching, Germany.\\
  $^{12}$NRC Herzberg Institute of Astrophysics, 5071 West Saanich
  Road, Victoria, V9E
  2E7, Canada.\\
}

\date{ACCEPTED SEPTEMBER 2011}

\begin{document}
\pagerange{\pageref{firstpage}--\pageref{lastpage}} \pubyear{2011}

\def\LaTeX{L\kern-.36em\raise.3ex\hbox{a}\kern-.15em
    T\kern-.1667em\lower.7ex\hbox{E}\kern-.125emX}

\maketitle
\label{firstpage}

\begin{abstract}

  We present weak lensing data from the HST/STAGES survey to study the
  three-dimensional spatial distribution of matter and galaxies in the
  Abell 901/902 supercluster complex.  {Our method improves over
    the existing 3D lensing mapping techniques by calibrating and
    removing redshift bias and accounting for the effects of the
    radial elongation of 3D structures.} {We also include the
    first detailed noise analysis of a 3D lensing map, showing that
    even with deep HST quality data, only the most massive structures,
    for example \mbox{$M_{200}\gtrsim10^{15} M_\odot \rm{h}^{-1}$} at
    \mbox{$z\sim 0.8$}, can be resolved in 3D with any reasonable
    redshift accuracy ($\Delta z\approx0.15$).}  We compare the
  lensing map to the stellar mass distribution and find luminous
  counterparts for all mass peaks detected with a peak significance
  \mbox{$>3\sigma$}.  We see structures in and behind the $z=0.165$
  foreground supercluster, finding structure directly behind the A901b
  cluster at \mbox{$z\sim0.6$} and also behind the SW group at
  \mbox{$z\sim0.7$}. {This 3D structure viewed in projection
    has no significant impact on recent mass estimates of A901b or the
    SW group components SWa and SWb.}

\end{abstract}

\begin{keywords}
dark matter - large-scale structure of Universe - gravitational
lensing
\end{keywords}


\section{Introduction}

Our current understanding of cosmology describes an intricate web of
dark matter spanning the Universe dictating when and where galaxies
form.  These large-scale structures of dark matter have long been
simulated in large N-body computations \citep{Millennium}, but due to
the nature of dark matter, linking this feature of dark matter theory
directly to observations has been a challenging task.

Historically, the presence of dark matter as an essential ingredient
of the standard model of cosmology was deduced indirectly from the
rotation of galaxies or the stability of galaxy clusters as the
largest known cosmological objects. {Nowadays, the dark matter
  density is inferred from the break-scale in the galaxy power
  spectrum, and by combining fluctuations in the cosmic microwave
  background, the luminosity-density relation of supernovae Type Ia,
  the Lyman-$\alpha$ forest, or the bulk motions of galaxies.}  For an
overview, see for example \citet{2003moco.book.....D} and
\citet{1999coph.book.....P} and references therein.  Nevertheless, a
direct detection of dark matter particles as such is still an open
issue.

Weak gravitational lensing offers a unique way to chart dark matter
structures in the Universe as it is sensitive to all forms of matter.
Weak lensing has been used to map the dark matter in galaxy clusters
\citep[see for example][]{Clowe06} with high resolution
reconstructions recovered for the most massive strong lensing clusters
\citep[see for example][]{Bradac06}.  Several lensing studies have
also mapped the projected surface mass density over degree scale
fields \citep{GavazziSoucail,Schirmer07,Kubo09} to identify
{shear selected} groups and clusters.  The minimum mass scale
that can be identified is limited only by the intrinsic ellipticity
noise in the lensing analysis {and projection effects}.  Using
a higher number density of galaxies in the shear measurement reduces
this noise, and for this reason the Hubble Space Telescope (HST) is
truly unique for this area of research.  {Compared to
  ground-based observatories, observations from space more than triple
  the number density of galaxies that can be resolved with a
  significantly enhanced imaging quality, permitting high resolution
  reconstructions of dark matter in the field \citep{MasseyNat,
    Heymans08} and the study of lenses at higher redshift.}

The lensing distortion experienced by a source galaxy depends on the
total projected surface mass density along the line of sight (l.o.s.)
and a geometrical factor that increases with source distance.  The
highest redshift galaxies therefore experience, on average, the
strongest shear distortions.  This redshift dependence can be used to
recover the full three-dimensional gravitational potential of the
matter density \citep{taylor01,bacontay} and has been applied to the
\mbox{COMBO-17} survey in \citet{Taylor04} and the \mbox{COSMOS}
survey in \citet{MasseyNat}.  All these 3D mass reconstruction
methods however require the use of a prior based on the expected mean
growth of matter density fluctuations.  Without the inclusion of such
a prior, \citet{hukeeton02} show that a Wiener filter approach is
unable to reasonably constrain the radial matter distribution, even
for densely sampled space-based quality lensing data.

An advancement of the method of \citet{hukeeton02} is presented in
\citet{STH09}, hereafter STH09, which details a non-parametric
reconstruction method that directly reconstructs the 3D matter
density distribution.  This is in contrast to previous methods which
have mapped the gravitational potential.  {In broad terms one can
  view the gravitational potential as a heavily smoothed realisation
  of the density field.  As an example, a density map of a single
  point mass will have a broad corresponding gravitational potential
  map that radially decays as $1/r$ in 3D (as $\ln{r}$ in a 2D on
  the sky projection) from the position of the source.  Directly
  reconstructing the density field allows one to see the sources that
  produce the smoother gravitational potential through Poisson's
  equation and therefore this is ultimately what we wish to do.
  Historically, the 3D potential field has been mapped because the
  signal-to-noise was higher. Applying a Laplacian to a noisy
  potential field appeared unfeasible on cluster scales. However, the
  work of STH09 showed that the Wiener filter method could also be
  applied in this regime to recover the matter-density field. Since
  the matter-density is the quantity we are interested in, and because
  the mass sheet degeneracy leads to unconstrained quadratic terms in
  the potential field, we now focus purely on the reconstruction of
  the 3D matter-density field. Since this is the first application of
  this method, we shall pay close attention to the changes in
  signal-to-noise.}

{Another new, computionally faster singular value decomposition
  technique has recently been presented in
  \citet{2011ApJ...727..118V}.  This paper proposes to use Eigenmodes
  in the reconstructed 3D mass density field, rejecting modes above a
  given noise level, thereby achieving a radial smoothing without a
  Wiener filter. On a deeper level, this is what a Wiener filter is
  doing as well, as it suppresses density modes according to their
  signal-to-noise, for which the prior is used to determine the
  expected signal in the reconstruction. }

The Wiener filter employed by STH09 uses a $\Lambda$CDM prior to
optimally smooth and reconstruct the 3D map.  The smoothing does,
however, have an unwanted impact as it (1) spreads out structures in
radial (redshift) direction, (2) shifts on average the centre of
structures to different redshifts (here-after referred to as $z$-shift
bias) and (3) scales the overall amplitude of matter density
fluctuations. {We note here that while the radial spread and
  scaling of the amplitude of recovered perturbations was noted and
  detected in the 3D potential mapping approach, shifts in the
  position of structure were first pointed out in STH09 and therefore
  not accounted for in these studies.}  The information about how the
true matter distribution is smoothed in the resulting map is clearly
defined by a radial point spread function (p.s.f).  This radial p.s.f
needs to be taken into full account for any correct interpretation of
the map.  In addition, as also outlined in STH09, the redshift
resolution of all 3D lensing reconstruction will be low owing to the
weak response of lensing to variations in lens redshift encoded in the
broad lensing kernel.

The STH09 method has been optimised for observational data as it
utilises the lensing information from all galaxies, whether they are
with or without a photometric redshift estimate.  It also has inbuilt
tests for the effects of systematic errors in the lensing measurement,
using an E/B mode decomposition.  Crucially it also produces a 3D
noise map so the significance of structures in the resulting maps can
be assessed.

In this paper, we present the first application of the 3D matter
density non-parametric reconstruction method, analysing the \mbox{HST
  STAGES} survey of the Abell 901/902 supercluster field, hereafter
A901/902.  We choose this field to test this method as it contains
four known massive structures at $z \sim 0.165$ detected at high
significance in a 2D map \citep{Heymans08} and a fifth known
structure in projection at $z \sim 0.46$ \citep{Taylor04}.  In
addition, STAGES is one of the largest HST surveys with one of the
highest quality lensing data sets for this type of analysis.

The 3D mass map is presented as a raw Wiener 3D reconstruction that
reveals the expected low redshift resolution, based on our knowledge
of the radial p.s.f and lensing kernel.  Assuming the local maxima
along each l.o.s. are a good estimate of the position of the peak
density, we present 3D visualisations of the dark matter density in
this field.  Highlighting the most likely positions of structures in
this way produces remarkable images that can be compared to other
published 3D lensing reconstructions.  However we voice strong
caution in using any of these visualisations for scientific purposes
as they present a very noisy estimate of the density field and contain
no error information. Moreover, structure along the same l.o.s. can
overlap and thereby merge in the Wiener map to one single structure at
a combined average redshift.

For scientific purposes, we propose the use of raw Wiener
reconstructions. We identify interesting l.o.s. across the field from
a 3D cross-correlation analysis of mass and light and then fully take
into account the radial biasing and smoothing of the Wiener
reconstruction to determine each lens redshift probability
distribution. We then compare this probability distribution to
over-densities of stellar mass along the same line l.o.s. and discuss
our findings for the 3D matter distribution in the STAGES field.
  
The paper is set out as follows. In Sect.~\ref{sec:method} we review
the STAGES data set and the 3D reconstruction method and present our
results in Sect.~\ref{sec:results}.  We discuss our findings and
conclude in Sect.~\ref{sec:discuss}.  As this analysis places new
requirements on the accuracy of the weak lensing observations, the
Appendix details our tests of the observational analysis that ensure
the results presented in this paper are robust.

\section{3D mass reconstruction of STAGES} 
\label{sec:method}

In this section, we focus on the application of the STH09
non-parametric 3D matter density reconstruction method to the STAGES
A901/902 supercluster field, referring the reader to STH09 for details
on the full theory that underpins this method.  This field has been
the subject of several weak lensing analyses
\citep{Gray02,Taylor04,Schirmer07,Heymans08}, investigating different
aspects of the dark matter environment of this dense field.  The most
recent analyses have focused on data from the Space Telescope A901/902
Galaxy Evolution Survey \citep[STAGES,][]{Gray09}.  In this analysis,
weak lensing distortions have been measured for over $60,000$ galaxies
\citep[$\sim 65$ galaxies per square arcmin;][]{Heymans08} of which
$\sim 15\%$ of the sample have accurate photometric redshift and
stellar mass estimates from the \mbox{COMBO-17} survey \citep{Wolf04}.
Lensed galaxies without photometric redshifts estimates are
incorporated using magnitude bins for which the redshift distribution
is estimated using \citet{Schrabback07}.

The 3D density reconstruction presented in this paper is the most
complex lensing analysis of these data and places new demands on the
accuracy and reliability of the weak lensing shear measurement and the
accuracy of the photometric redshifts.  In the Appendix, we therefore
document the steps taken to confirm the reliability of the weak
lensing shear catalogue that forms the basis of the 3D
reconstruction.

At the very end of this section, in subsection \ref{sect:execsummary},
we give an executive summary of the main results to highlight the new
results of this section.

\subsection{3D mapping in a nutshell} 

In weak gravitational lensing, gravity-induced shape distortions of
galaxy images are linearly related to the projected matter
distribution along a line of sight (l.o.s.) direction $\vec{\theta}$.
{Let us start with a simple case in which we bin the complex ellipticities of \emph{all} sources in a full shear catalogue onto the same regular grid. The grid covers the field-of-view of the survey, the mean ellipticity of the $i$th grid cell is denoted by $\epsilon(\vec{\theta}_i)$; $\vec{\theta}_i$ is the l.o.s. direction of the $i$th cell. We arrange the binned ellipticities as one vector
  \begin{equation}
    \vec{\epsilon}=
    \left(\epsilon(\vec{\theta}_1),
      \epsilon(\vec{\theta}_2),\ldots\right)^{\rm t}\;.
  \end{equation}

  In the working regime of weak gravitational lensing, the convergence
  $\kappa(\vec{\theta})$ of light bundles is the projection of the 3D
  matter density fluctuations $\delta(\vec{r}_\perp,\chi)$, where
  $\vec{r}_\perp$ is a comoving distance vector perpendicular to a
  fiducial l.o.s., and $\chi$ is the comoving radial distance
  \citep{2001PhR...340..291B}:
  \begin{equation}
    \label{eq:kappa}
    \kappa(\vec{\theta})=
    \frac{3H_0^2\Omega_{\rm m}}{2c^2}
    \int_0^\infty\d\chi\frac{\overline{W}(\chi)f_{\rm k}(\chi)}{a(\chi)}
    \delta\left(f_{\rm k}(\chi)\vec{\theta},\chi\right)\;.
  \end{equation}
  Here $c/H_0$ is the Hubble radius, $f_{\rm k}(\chi)$ the angular
  diameter distance of $\chi$, $\Omega_{\rm m}$ the mass density
  parameter and $a(\chi)$ the scale factor at distance $\chi$. The
  function $\overline{W}(\chi)$ expresses the lensing kernel averaged
  for a distribution of sources in $\chi$:
  \begin{equation}
    \overline{W}(\chi)=
    \int_\chi^\infty\d\chi^\prime\frac{f_{\rm
        k}(\chi^\prime-\chi)}{f_{\rm k}(\chi^\prime)}p_\chi(\chi^\prime)\;;
  \end{equation}
  $p_\chi(\chi)$ is the probability density of source distances.

  As Ansatz, we chop the matter distribution $\delta$ into a set of
  $N_{\rm lp}$ matter slices or lens planes. Within each matter slice,
  $\delta$ is assumed to be constant along a l.o.s. $\vec{\theta}$, so
  that the $j$th matter slice is represented by
  $\delta^{(j)}(\vec{\theta}_i)$, the average $\delta$ in direction
  $\vec{\theta}_i$ (lens plane), and the width
  $\Delta\chi_j=\chi_{j+1}-\chi_j$ of the slice; $\chi_j$ defines the
  boundaries of adjacent slices. By means of
  \begin{equation}
    \label{eq:deltadef}
    \vec{\delta}^{(j)}=
    \left(\delta^{(j)}(\vec{\theta}_1),\delta^{(j)}(\vec{\theta}_2),\ldots\right)^{\rm t}
  \end{equation}
  we arrange the grid values of $\delta$ of every slice as a
  vector. Note that we assume a flat sky where every lens
  plane is a tangential plane to the celestial sphere in the direction
  of the fiducial l.o.s..

  Likewise, we arrange the $\kappa(\vec{\theta})$ into a vector:
  \begin{equation}
    \vec{\kappa}=    
    \left(\kappa(\vec{\theta}_1),\kappa(\vec{\theta}_2),\ldots\right)^{\rm t}\;.
  \end{equation}
  With this Ansatz, Eq. \Ref{eq:kappa} can now be written as linear
  projection of the $\delta$-slices into the $\kappa$ grid:
  \begin{equation}
    \label{eq:kappa2}
    \vec{\kappa}=
    \sum_{i=1}^{N_{\rm lp}}Q_i\vec{\delta}^{(i)}\equiv\mat{Q}\vec{\delta}\;,
  \end{equation}
  where
  \begin{equation}
    Q_i=\frac{3H_0^2\Omega_{\rm m}}{2c^2}
    \int_{\chi_i}^{\chi_{i+1}}\d\chi\frac{\overline{W}(\chi)f_{\rm k}(\chi)}{a(\chi)}\;.
  \end{equation}
  In the last step of Eq. \Ref{eq:kappa2}, we denoted for convenience
  the weighed sum of $\vec{\delta}^{(i)}$'s by $\mat{Q}\vec{\delta}$.

  In order to relate the convergence on the grid to the binned
  ellipticities of our survey, we have to transform convergence to
  shear
  \begin{equation}
    \vec{\gamma}=
    \left(\gamma(\vec{\theta}_1),\gamma(\vec{\theta}_2),\ldots\right)^{\rm t}
  \end{equation}
  in the next step. On a flat sky, the transformation is expressed by
  the well-known \citet{1993ApJ...404..441K} relation, which is
  written down for a regular grid here \citep{hukeeton02}:
  \begin{equation}
    \vec{\gamma}=
    \mat{P}_{\gamma\kappa}\mat{Q}\vec{\delta}\;,
  \end{equation}
  where we have for the convergence-to-shear transformation matrix
  \begin{equation}
    [\mat{P}_{\gamma\kappa}]_{ij}=
    \left\{\begin{array}{ll}
        -\frac{A}{\pi[\theta_{ij}^\ast]^2} & i\ne j\\
        0 & i=j
        \end{array}\right.
  \end{equation}
  with $\theta_{ij}=\vec{\theta}_i-\vec{\theta}_j$ and $A$ being the
  solid angle of a grid pixel. Note that for the 2D vector
  $\vec{\theta}$ we are using the commonly employed complex notation.

  So far, nothing has been said about the noise,
  \begin{equation}
    \vec{n}=\left(n(\vec{\theta}_1),n(\vec{\theta}_2),\ldots\right)^{\rm
      t}\;,
  \end{equation}
  which is necessarily included in $\vec{\epsilon}$ owing to the
  intrinsic shape of the sources and the non-uniform distribution of
  sources resulting in occasional cases of grid cells with zero
  sources (for example in a masked region):
  \begin{equation}
    \vec{\epsilon}=\vec{\gamma}+\vec{n}=
    \mat{P}_{\gamma\kappa}\mat{Q}\vec{\delta}+\vec{n}\;.
  \end{equation}
  Of $\vec{n}$ only the statistical properties are known: it is
  assumed to be uncorrelated to $\vec{\delta}$, has a vanishing
  statistical average and the covariance
  $\mat{N}=\ave{\vec{n}\vec{n}^\dagger}$. In the case of missing
  sources, the noise $n(\vec{\theta}_i)$ associated with the $i$th
  grid cell is infinite.}

In 3D lensing, we have a survey that can be split into various
subsamples of sources with distinct and known redshift
distributions. Now each source subsample provides a different grid
$\vec{\epsilon}_i$, which every time is a projection of the same
matter cube $\vec{\delta}$ but with a different set of weights
$\mat{Q}_i$ and noise grids $\vec{n}_i$. Therefore, employing a
compact notation, we now have \begin{eqnarray}
  \label{eq:mapproblem}
  \nonumber
  \vec{\epsilon}&\equiv&
  \left(\vec{\epsilon}_1,\vec{\epsilon}_2,\ldots\right)\\
  \nonumber
  &=&
 \mat{P}_{\gamma\kappa}\left(\mat{Q}_1\vec{\delta},\mat{Q}_2\vec{\delta},\ldots\right)+
  \left(\vec{n}_1,\vec{n}_2,\ldots\right)\\
  &\equiv&
  \mat{P}_{\gamma\kappa}\mat{Q}\vec{\delta}+\vec{n}\;.
\end{eqnarray}

{
  The map making algorithm utilises this relation to define an
  estimator for $\vec{\delta}$ given $\vec{\epsilon}$ and the noise
  properties $\mat{N}$. Although Eq. \ref{eq:mapproblem} has no exact
  solution due to the unknown noise component $\vec{n}$ and the
  non-invertible $\mat{P}_{\gamma\kappa}$ (mass sheet degeneracy), one
  can find an optimised solution, i.e. a linear filter $\mat{H}$ of
  $\vec{\epsilon}$, that minimises the average deviation between the
  estimated minimum variance $\vec{\delta}_{\rm
    mv}=\mat{H}\vec{\epsilon}$ and the true $\vec{\delta}$, i.e.
\begin{equation}
  \ave{\parallel\vec{\delta}_{\rm mv}-\vec{\delta}\parallel^2}={\rm min.}
\end{equation}
(By $\parallel\vec{x}\parallel^2=\vec{x}^{\rm t}\vec{x}$ we mean the
Euclidean norm of $\vec{x}$.)

The solution to this optimisation problem is in our
case \citep[e.g.][]{1995ApJ...449..446Z}:
\begin{equation}
  \mat{H}=
  \left[\alpha\mat{1}+\mat{S}\mat{R}^\dagger\mat{N}^{-1}\mat{R}\right]^{-1}
  \mat{S}\mat{R}^\dagger\mat{N}^{-1}\;,
\end{equation}
where $\mat{R}\equiv\mat{P}_{\gamma\kappa}\mat{Q}$, and $\alpha$ is an
additional constant, usually between $\alpha\in[0,1]$, that allows to
change the signal-to-noise in the resulting map by regulating the
smoothing of the filter \citep[Saskatoon
filter;][]{1997ApJ...480L..87T}.  The signal covariance
$\mat{S}\equiv\ave{\vec{\epsilon}\vec{\epsilon}^\dagger}-\mat{N}$ has
to be given to the filter as prior and is computed for a particular
fiducial cosmological model. The practical challenge is the
computation of $\vec{\delta}_{\rm mv}=\mat{H}\vec{\epsilon}$ within a
reasonable time, see the Appendix of STH09 for details.}

To evade the problem of the mass-sheet degeneracy, we restrict
ourselves to solutions for which the average of $\vec{\delta}_{\rm
  mv}$ vanishes for every redshift slice individually.

\subsection{Reconstruction prior and grid parameters} 
\label{sec:alpha}

The prior can be implemented in the form of a radial or a transverse
filter; a transverse filter varies as a function of angular wave number
$\ell$ (STH09).
{For $\mat{S}$, we choose a radial Wiener filter 
\begin{equation}
  \ave{\delta^{(i)}(\vec{\theta}_j)\delta^{(k)}(\vec{\theta}_l)}
  =
  \frac{\delta^{\rm K}_{jl}\delta^{\rm K}_{ik}}{2\pi}\int_0^\infty
  \!\!\!\!\d\ell\,\ell\,
  P_\delta^{(i)}(\ell)|F(\Theta_{\rm s}\ell)|^2J_0(\ell\theta)\;,
\end{equation}
where $F(x)=2x^{-1}J_1(x)$ is the Fourier transform of a top-hat
window, $J_n(x)$ a Bessel function of first kind and
$P_\delta^{(i)}(\ell)$ is the power of matter fluctuations, projected
along the width of the $i$th matter slice. We approximate our square
grid pixels by circular pixels of identical area $A=\pi\Theta_{\rm
  s}^2$.}

As a fiducial matter power spectrum we use the approximation of
\citet{Smith03} and Limber's equation for converting 3D to 2D power
on the sky. The fiducial cosmology is a $\Lambda$CDM model (adiabatic
fluctuations) with \mbox{$\Omega_{\rm m}=0.24$}, \mbox{$\Omega_{\rm
    b}=0.0416$}, \mbox{$\Omega_{\Lambda}=1-\Omega_{\rm m}$} and
\mbox{$H_{0}=h\,100\,{\rm km}\,{\rm s}^{-1}\,{\rm Mpc}^{-1}$} with
\mbox{$h=0.732$}. The normalisation of the matter fluctuations within
a sphere of radius $8\,h^{-1}\rm Mpc$ at redshift zero is assumed to
be $\sigma_8=0.76$. For the spectral index of the primordial matter
power spectrum we use $n_{\rm s}=0.96$. These values are consistent
with the third-year WMAP results \citep{2007ApJS..170..377S}.

The smoothing of the Wiener filter can be regulated by rescaling the
noise covariance of the data via $\alpha$: For $\alpha=0$ Wiener
filtering is switched off, i.e. the prior $\mat{S}$ cancels out at the
expense of the lowest signal-to-noise of the reconstruction but for
the benefit of an unbiased estimate of $\vec{\delta}$.  For $\alpha=1$
the Wiener filter boosts the signal-to-noise to a maximum at the
expense of heavy smoothing in the radial direction along with a strong
$z$-shift bias and loss in redshift resolution.  We find
$\alpha=10^{-3}$ provides a good balance to maximise signal-to-noise
and minimise bias, which we fully account for below. Note, a radial
prior does not produce transverse smoothing so an additional
transverse smoothing is required for every lens plane. For
post-reconstruction smoothing we use a Gaussian kernel of $\sigma_{\rm
  rms}=40$ arcsec, which is the size of several grid pixels.

The number of matter lens planes used here is $N_{\rm lp}=22$,
covering a range from \mbox{$z=0\ldots1.15$} with an associated slice
width of \mbox{$\Delta z=0.05$}. There is one exception for the
closest plane which has \mbox{$\Delta z=0.1$}.
{The width of the slices is finer than the radial resolution we can
  possibly hope for. Nevertheless, the large number of lens planes was
  chosen to avoid numerical instabilities in the reconstruction.}  The
lens planes are gridded with an angular resolution of
\mbox{$128\times128\,\rm pixel^2$}, one pixel has a square size of
$16.34$ arcsec.
{The vector $\vec{\delta}$ hence has $128^2\times N_{\rm lp}\approx3.6\times10^5$ degrees of freedom.}

\subsection{The radial point spread function}
\label{sect:radialpsf}

As outlined in STH09, the reconstruction algorithm does not provide an
unbiased representation of the true radial matter distribution. The
bias can however be quantified.

Ignoring the noise in a particular reconstruction, on average the map
$\vec{\delta}_{\rm mv}$ will be a smoothed image of the true matter
distribution $\vec{\delta}$:
\begin{equation}
  \ave{\vec{\delta}_{\rm mv}}=
  \mat{H}\ave{\vec{\epsilon}}=
  \mat{H}\mat{P}_{\gamma\kappa}\mat{Q}\ave{\vec{\delta}+\vec{n}}
  =\mat{H}\mat{P}_{\gamma\kappa}\mat{Q}\vec{\delta}\;,
\end{equation}
where $\ave{\ldots}$ is the ensemble average over all source noise
realisations while keeping the actual matter distribution
$\vec{\delta}$ constant.  Hence the matrix
$\mat{B}=\mat{H}\mat{P}_{\gamma\kappa}\mat{Q}$ holds all information
on the smoothing involved in the reconstruction process. In the
optimal case, one would have $\mat{B}=\mat{1}$. Here, we do not have
$\mat{B}=\mat{1}$, though, but still $\mat{B}$ contains all
information needed for a quantitative assessment of the resulting map.

Unfortunately, as discussed in STH09, simply applying the inverse of
$\mat{B}$ to $\vec{\delta}_{\rm mv}$, even if $\mat{B}$ was regular,
would undo the Wiener filter, leaving behind a heavily oscillating
reconstruction with a significant noise level because low
signal-to-noise modes in the map would no longer be suppressed.

Here, the full matrix $\mat{B}$ has of the order of
$10^5\times10^5=10^{10}$ elements. It turns out, however, that in
practise most of its elements are negligible because the devised
Wiener filter smoothes the matter distribution only in the radial
direction.

Concretely, consider that we have just one point source overdensity
inside the entire 3D density field $\vec{\delta}$, represented by one
unity-valued grid pixel in direction $\vec{\theta}_i$ on the $j$th
lens plane. To lift the mass-sheet degeneracy, the algorithm assumes
that the density contrast averaged over \emph{every} lens plane
vanishes. Hence, to stay inside the space of considered
$\vec{\delta}$-configurations all other pixels of the $j$th lens plane
have to be set to $-(127\times128)^{-1}$ for our $(128\,\rm pixel)^2$
grids; pixels of all other lens planes $i\ne j$ are zero. We denote
this special density contrast configuration by $\vec{e}^{(j)}_i$. Any
arbitrary other configuration $\vec{\delta}$, with zero average
contrast for each lens plane, can be represented as weighed sum of
$\vec{e}^{(j)}_i$'s as $\vec{\delta}=\sum_{i,j}c_{ij}\vec{e}^{(j)}_i$
with coefficients $c_{ij}$. The reconstruction algorithm is linear so
that the resulting noise averaged Wiener map is a linear combination
of all $\mat{B}\vec{e}^{(j)}_i$, i.e.
\begin{equation}
  \mat{H}\ave{\mat{P}_{\gamma\kappa}\mat{Q}\vec{\delta}+\vec{n}}=
  \mat{B}\vec{\delta}+\mat{H}\ave{\vec{n}}=
  \sum_{i,j}c_{ij}\mat{B}\vec{e}^{(j)}_i\;;
\end{equation}
the vector $\vec{n}$ denotes the shear noise in a particular
realisation.

Essentially the original matter point source $\vec{e}^{(j)}_i$ is
smeared out in the l.o.s. direction $\vec{\theta}_i$, but not in the
transverse.  In \mbox{STAGES} the radial smearing will be roughly
unchanged when varying $\vec{\theta}_i$ inside the same lens plane
$i$. We checked this by computing $\mat{B}\vec{e}^{(j)}_i$ for a set
of $N_{\rm los}=10\times10$ l.o.s. directions for all lens planes
covering the inner $80\%$ of the field. Therefore, instead of keeping
the full $\mat{B}\vec{e}^{(j)}_i$ for all $\vec{e}^{(j)}_i$, we store
only the average l.o.s. profile of $\mat{B}\vec{e}^{(j)}_i$ along
$\vec{\theta}_i$ and average over several $\vec{\theta}_i$:
\begin{equation}
  \label{eq:radialpsf}
  \vec{p}^{(j)}=
  \frac{1}{N_{\rm los}}\sum_{i=1}^{N_{\rm los}}[\mat{B}\vec{e}^{(j)}_i,\vec{\theta}_i]\;,
\end{equation}
where $[\vec{A},\vec{\theta}_i]$ denotes the linear operation that
takes only elements from the 3D grid $\vec{A}$ that are along the
l.o.s. direction $\vec{\theta}_i$ and arranges them as an $N_{\rm lp}$
element vector in order of ascending lens plane redshifts. Note that
$\mat{B}\vec{e}^{(j)}_i$ is calculated in practise by applying the
reconstruction algorithm to
$\mat{P}_{\gamma\kappa}\mat{Q}\vec{e}^{(j)}_i$ so that in total
$N_{\rm los}^2\times N_{\rm lp}=2200$ individual reconstructions were
performed.

\begin{figure}
  \begin{center}
    \psfig{file=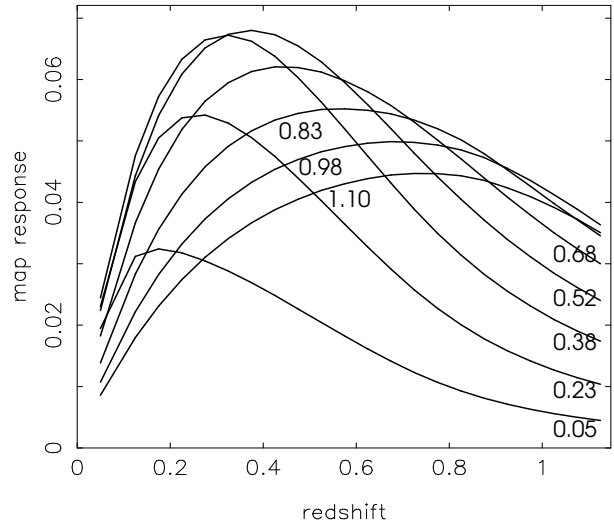,width=80mm,angle=0}
  \end{center}
  \caption{\label{fig:radpsf}
    {Average radial smearing for different lens plane redshifts, shown as small numbers attached to the profiles. Note that the local maximum of a p.s.f.  generally does not coincide with the lens plane redshift.}}
\end{figure}

Fig. \ref{fig:radpsf} shows the resulting radial p.s.f. for a selected
set of lens planes. Radially located peaks in the true matter
distribution will have a cigar-like appearance in the map due to the
evident width of the p.s.f. The spread will be reduced for better
signal-to-noise in the data, but will always be present to some degree
when a Wiener filter is employed.

\subsection{$Z$-shift bias}
\label{sec:zbias}

 \begin{figure}
    \begin{center}
      \psfig{file=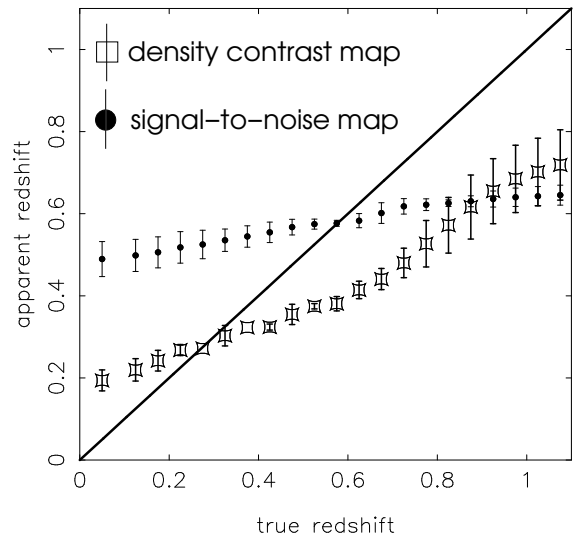,width=75mm,angle=0}
    \end{center}
    \caption{\label{fig:zshiftbias}
      {$Z$-shift bias of the radial filter used in the
        reconstruction. Owing to the Wiener filtering, structures do
        not appear at the correct redshift in the lens planes as
        indicated by the data points. Using this figure the $z$-shift
        bias in the reconstructed 3D map can be corrected for (see
        text). The error-bars denote the r.m.s.-variance of the bias
        over the field-of-view of \mbox{STAGES}. {\emph{Open squares}:
        bias in the density contrast map; \emph{filled points}: bias in the S/N map.}}}
  \end{figure}

  An obvious application of the Wiener map is to estimate the
  redshifts of 3D peaks by the local maximum of the cigar density
  profile {or to compare the redshift estimate to other maps of
    the galaxy number density or luminosity.} However, as can be seen
  from Fig. \ref{fig:radpsf}, generally the maximum of the p.s.f. does
  not coincide with the true redshift of structures, which we
  hereafter refer to as $z$-shift bias.  To determine the extent of
  $z$-shift bias we use Fig.~\ref{fig:radpsf} to determine where, on
  average a point source mass is recovered in the reconstruction.  The
  result is shown in Fig. \ref{fig:zshiftbias}, which displays the
  redshift of a p.s.f. local maximum (apparent redshift) and error in
  relation to the original point source redshift in
  $\vec{e}^{(j)}_i$. We use the data points in this figure to
  construct a new Wiener map free of the $z$-shift bias in the
  following way: We take a set of new lens planes with ``true
  redshifts'' and fill for every $\vec{\theta}_i$ these planes with
  values from the uncorrected Wiener map at the ``apparent redshift''
  given by Fig. \ref{fig:zshiftbias}.

\subsection{Quantifying noise and systematics}

{

  To assess the S/N of features inside the map, we divide map values
  by the variance observed in $10^3$ map noise realisations before
  $z$-shift bias correction. Noise maps are realised by randomly
  rotating the orientations of sources and rerunning the
  reconstruction algorithm.  During this process, we also average the
  noise covariance $\mat{N}$ of density pixels along the same l.o.s..

  Dividing by the noise variance inside a lens plane changes the shape
  of the p.s.f. and the $z$-shift bias, see open squares in
  Fig. \ref{fig:zshiftbias}. In the case of \mbox{STAGES}, the S/N-map
  shifts all p.s.f. peaks essentially to the middle of the map (in
  redshift), which can be explained by larger levels of noise variance
  at low and high redshifts. Due to the strong impact of the $z$-shift
  bias, the S/N-map is in essence thus only a 2D map stretched out
  over the entire redshift range.  For that reason, we will use it in
  the following only as a 2D projection to quote the maximum S/N
  expected in a l.o.s. direction.  }

To test the impact of systematic errors in the lensing shear
measurement, B-mode maps are produced. These are obtained by rotating
the source ellipticities by $45$-degrees followed by a repetition of
the reconstruction process. Ideally, these maps should be pure noise
maps as a gravitational tidal field is unable to generate a B-mode
distortion by shearing galaxy images. Instead, they arise from
systematics related to the source shape p.s.f. correction procedure,
the intrinsic clustering of sources \citep{SvWM02} and intrinsic
galaxy alignment correlations of physically close sources or
shape-shear-correlations \citep[][and references
therein]{Heymans06}. Moreover, as discussed in STH09, matter
fluctuations outside the field-of-view but close to the field-of-view
may be capable of generating large-scale B-modes within the
reconstruction volume. Therefore, an entirely vanishing level of
B-modes is never guaranteed per se. To partly suppress this effect, we
take a reconstruction field-of-view somewhat larger than the actual
coverage of the survey and later {remove} the area not covered by
observations. This allows {unknown} sources of shear to be
outside the field-of-view and to be properly accounted for in the
{model reconstruction.}

The peak and trough statistics in Sect. \ref{sect:noisepeaks} shows
that the B-mode map is consistent with noise realisations, whereas the
E-mode map is inconsistent with noise at high confidence as it should
be. This leads us to the conclusion that any systematic errors are low
and within our statistical errors.

\subsection{Redshift uncertainties}

The main drawback of the Wiener method is that the radial spread of
the p.s.f. hampers the ability to attach a redshift to a cigar-shaped
profile in the map.  This would be true even if a p.s.f. was very
narrowly peaked as the uncertainty of the estimate still needs to be
determined by accounting for the noise in the map, which is not
directly visible in the Wiener estimate for the 3D matter density
contrast.

What really determines the redshift accuracy is the ability to
distinguish a radial profile $\vec{p}^{(i)}$ from $\vec{p}^{(j)}$,
Eq. \Ref{eq:radialpsf}, despite of the noise superimposed onto the
average profile in a concrete realisation.

To get a handle on the principle redshift accuracy, let us consider a
simple case where the map density profile in the l.o.s. direction is
dominated by one single structure on the $i$th lens plane,
i.e. $\vec{\delta}^{(i)}=\delta_0\vec{p}^{(i)}+\vec{n}$; $\delta_0$ is
the grid pixel density contrast and $\vec{n}$ the noise vector in that
particular realisation.  To pin down the redshift of the structure
from the noisy vector $\vec{\delta}^{(i)}$ means whether we can
statistically distinguish, given the data, a model (a) in which the
mass is presumed to be on lens plane $j$ from the correct model (b)
where the mass is on lens plane $i$. Accepting model (a), we would
estimate the density $\delta^\prime_0=\lambda\delta_0$ at redshift
$z_j$ by minimising
\begin{equation}
  \chi^2(\lambda)=
  \delta^2_0\left(\lambda\vec{p}^{(j)}-\vec{\delta}^{(i)}\right)^{\rm
    t}
  \mat{N}^{-1}
  \left(\lambda\vec{p}^{(j)}-\vec{\delta}^{(i)}\right)\;.
\end{equation}
We are assuming a Gaussian noise model with covariance
$\mat{N}=\ave{\vec{n}\vec{n}^{\rm t}}$ here. Since the source noise is
roughly Gaussian and the Wiener filter linear this is a valid
assumption here. The minimum solution for the previous equation is
\begin{equation}
  \lambda_{\rm min}=
  \frac{[\vec{p}^{(j)}]^{\rm t}\mat{N}^{-1}\vec{\delta}^{(i)}}
  {[\vec{p}^{(j)}]^{\rm t}\mat{N}^{-1}\vec{p}^{(j)}}\;,
\end{equation}
so that the $\chi^2$ of the best-fit of model (a) has 
\begin{equation}
  \chi^2_{ij}=
  \left(\lambda_{\rm min}\delta_0\vec{p}^{(j)}-\vec{\delta}\right)^{\rm t}
  \mat{N}^{-1}\left(\lambda_{\rm min}\delta_0\vec{p}^{(j)}-\vec{\delta}\right)\;,
\end{equation}
which on average for all realisations $\vec{n}$ is
\begin{eqnarray}
  \ave{\chi^2_{ij}}&=&\Delta\chi^2_{ij}+N_{\rm lp}\;,\\
  \Delta\chi^2_{ij}&=&\delta^2_0\left(\overline{\lambda}_{\rm min}\vec{p}^{(j)}-\vec{p}^{(i)}\right)^{\rm t}
  \mat{N}^{-1}\left(\overline{\lambda}_{\rm
      min}\vec{p}^{(j)}-\vec{p}^{(j)}\right)\;,\\
  \overline{\lambda}_{\rm min}&=&
  \frac{[\vec{p}^{(j)}]^{\rm t}\mat{N}^{-1}\vec{p}^{(i)}}
  {[\vec{p}^{(j)}]^{\rm t}\mat{N}^{-1}\vec{p}^{(j)}}\;.
\end{eqnarray}
Therefore, to distinguish model (a) and model (b), we have to decide
when $\ave{\chi^2_{ij}}$ is inconsistent with a $\chi^2$-distribution
with $N_{\rm lp}$ degrees of freedom, i.e. the $\chi^2$-distribution
of model (b). Since model (b) has $\bar{\lambda}_{\rm min}=1$ and
$\ave{\chi_{ij}^2}=N_{\rm lp}$, the difference in $\ave{\chi^2}$ is
simply the above $\Delta\chi^2_{ij}$.
    
To get fiducial values of $\Delta\chi_{ij}^2$ for \mbox{STAGES}, we
model $\delta_0$ as the centre of a singular isothermal sphere (SIS)
with mass profile \citep[e.g.][]{2001A&A...378..361B}
\begin{equation}
  \delta_{\rm sis}(\theta)=
  \frac{4\pi}{3}\frac{c^2a(\chi_{\rm h})}{H_0^2\Omega_{\rm m}}
  \left(\frac{\sigma_{\rm v}}{c}\right)^2\frac{1}{f_{\rm
      k}(\chi_{\rm h})\theta}\;,
\end{equation}
where $\chi_{\rm h}$ is the SIS comoving distance and $\sigma^2_{\rm
  v}$ the velocity dispersion of the SIS. The SIS is smoothed it about
the centre with a Gaussian kernel of size $\sigma_{\rm rms}=40\,\rm
arcsec$ to the Wiener map transverse smoothing scale:
\begin{equation}
  \label{eq:siscentral}
  \delta_0(\chi_{\rm h})=
  \frac{(2\pi)^{3/2}}{3}
  \frac{c^2a(\chi_{\rm h})}{H_0^2\Omega_{\rm m}}
  \left(\frac{\sigma_{\rm v}}{c}\right)^2
  \frac{1}{\Delta\chi_{\rm h}f_{\rm  k}(\chi_{\rm h})\sigma_{\rm rms}}\;.
\end{equation}
Note that we also have to divide the smoothed $\delta_{\rm sis}$ (SIS
density contrast integrated along the l.o.s.)  by the width
$\Delta\chi_{\rm h}$ of the matter slice it resides in, to comply with
the definition of the mean density contrast in our mapping algorithm,
Eq. \Ref{eq:kappa2}.

\begin{figure}
  \begin{center}
    \psfig{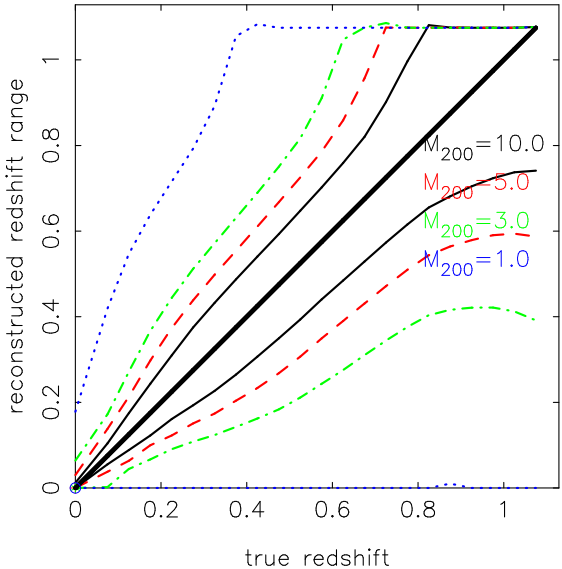}
  \end{center}
  \caption{\label{fig:zresolution}
    {Expected uncertainties of redshifts of structures with a SIS-like density profile as function of $M_{200}$ mass (units of $10^{14}\msol\,h^{-1}$). The thick solid black line shows the average redshift (true redshift: $x$-axis; apparent redshift: $y$-axis) of a SIS (no $z$-shift bias). The upper and lower lines bracket the $68\%$ confidence region for the scatter of redshifts inside the map. The width of the confidence interval depends on the SIS mass and its true redshift.}}
  \end{figure}

  Based on the Ansatz outlined above, Fig. \ref{fig:zresolution}
  displays the expected accuracy with which redshifts of structures
  can be determined from a Wiener filtered 3D map. In this plot,
  masses are defined as SIS-$M_{200}$ masses, i.e. the mass contained
  inside a radius $r_{200}$ within which the mean density equals $200$
  times the critical density:
  \begin{eqnarray}
    \label{eq:sism200}
    M_{200}&=&\frac{2^{3/2}\sigma_{\rm v}^3}{10GH(z)}\\
    \nonumber
    &\approx&
    6.58\times10^{14}\msol h^{-1}\,\left(\frac{\sigma_{\rm
          v}}{10^3\,{\rm kms^{-1}}}\right)^3
    \left(\frac{H(z)}{H_0}\right)^{-1}\;.
  \end{eqnarray}
  For comparison, the two most massive clusters in the field, A901a
  and A901b, have approximately a mass of
  $M_{200}\approx2\times10^{14}\msol h^{-1}$ \citep{Heymans08}.  The
  figure adopts the average radial p.s.f. of \mbox{STAGES}.  For a
  \mbox{STAGES}-like survey we find the redshift of a massive galaxy
  cluster of \mbox{$M_{200}\sim10^{15}\msol\,h^{-1}$} could still be
  measured with accuracy of \mbox{$\Delta z\approx\pm0.08$}
  (\mbox{$\pm0.15$}) at \mbox{$z\sim0.6$}
  (\mbox{$0.8$}), whereas a cluster of mass with
    \mbox{$M_{200}\sim3\times10^{14}\msol\,h^{-1}$} has \mbox{$\Delta
      z\approx0.05$} at \mbox{$z\sim0.2$} but already \mbox{$\Delta
      z\approx0.15$} at \mbox{$z\sim0.6$}. For masses smaller or equal
    than \mbox{$M_{200}\sim10^{14}\msol\,h^{-1}$}, the accuracy
    quickly deteriorates with redshift, at most imposing upper limits
    on the redshift. {This lack of 3D resolution for lower mass
      halos poses the natural question of how useful a 3D lensing
      reconstruction is, which we discuss in
      Section~\ref{sec:discuss}.}

  \begin{figure}
    \begin{center}
      \psfig{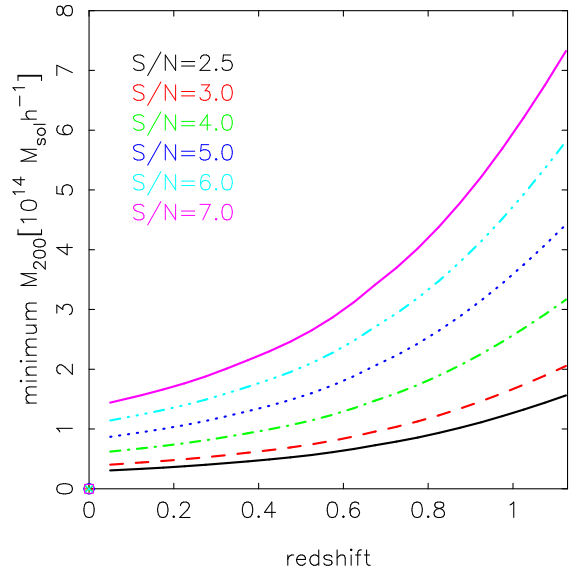}
    \end{center}
    \caption{\label{fig:minmass}{The minimum $M_{200}$ mass required to have a detection of a SIS located at a certain redshift ($x$-axis) above a certain significance level (different lines) \emph{anywhere} in the Wiener map of \mbox{STAGES}.}}
  \end{figure}

\subsection{Detection limit for mass concentrations}

{
  In order to gauge the mass detection limit of the \mbox{STAGES}
  Wiener map, we have produced Fig. \ref{fig:minmass}. For this plot,
  we place the central pixel $\delta_0(z_i)$ of a SIS at lens plane
  redshift $z_i$ and determine the maximum S/N ratio,
  \begin{equation}
    R_{\rm sn}={\rm max}
    \left\{
      \frac{[\delta_0(z_i)\vec{p}^{(i)}]_j}{\sqrt{[\mat{N}]_{jj}}}
      \big|j\in\{1,\ldots,N_{\rm lp}\}\right\}\;,
  \end{equation}
  along the SIS l.o.s. in the Wiener map; \mbox{STAGES} survey
  parameters are assumed. For a given $z_i$, $M_{200}$ is changed
  until $R_{\rm sn}$ reaches a desired S/N limit. This limit is easily
  found due to the scaling \mbox{$M_{200}\propto R_{\rm sn}^{1.5}$},
  which follows from Eqs. \Ref{eq:sism200} and \Ref{eq:siscentral}.

  For A901a/b at $z=0.165$ we expect a S/N between $6-7\sigma$ in the
  Wiener map, which indeed is found.  We can also see from this plot
  how well suited the STAGE field is for this type of analysis as we
  are primarily sensitive to structure at low and moderate redshift
  for $M_{200}\gtrsim10^{14}\msol h^{-1}$.
  To reconstruct mass at higher redshift requires a significant
  structure to lie in the field.}

\subsection{Product Map Mass and Stellar Mass comparison method}
\label{sect:productmap}

For a further analysis, we will compare the lensing mass map to the
distribution of stellar mass inside the data cube. The stellar mass
map is derived directly from the COMBO-17 galaxy catalogue, which is
binned onto the 3D grid used for the mass reconstruction.  We compare
the two 3D gridded distributions on a pixel-by-pixel basis with the
aim to find local matches and discrepancies between the light and
mass distribution.

For the mass-light comparison, we restrict the light map to a selected
redshift range:
\begin{equation}
  \hat{\vec{\delta}}_{\rm l}(\vec{\theta},k,l)=
  \sum_{i=k}^{l}\vec{e}_i\delta^{(i)}_{\rm
    l}\;,
\end{equation}
where the indices $l,k=1\ldots N_{\rm lp}$, $l\ge k$ choose the
redshift range in question; $\vec{e}_i$ denotes a vector with $N_{\rm
  lp}$ elements being zero everywhere except for the $i$th
element. The range is $z\in[z_k,z_l]$, where $z_i$ is the redshift of
the $i$th lens plane.

We understand the lensing map as a convolved underlying matter
distribution,
\begin{equation}
  \hat{\vec{\delta}}_{\rm m}=
  \sum_{i=1}^{N_{\rm lp}}\vec{p}^{(i)}\delta^{(i)}_{\rm
    m}\equiv\mat{P}\vec{\delta}_{\rm m}\;.
\end{equation}
The convolution of the mass map is an unavoidable by-product of the
Wiener filter reconstruction. Therefore, in contrast to the light map,
for the mass map we do not have the choice of selecting a certain
redshift range. We therefore cross-correlate stellar mass with the
full lensing mass map. This is reasonable because distinct lens planes
are well separated enough in physical distance to have no relevant
astrophysical correlation, i.e. for $i\ne j$
\begin{equation}
  \Ave{\delta_{\rm m}^{(i)}\delta_{\rm  l}^{(j)}}\approx0\;.
\end{equation}

Our analysis tool is a product map where the product is taken of the
stellar mass and total mass density profiles in radial direction:
\begin{equation}
  \label{eq:product}
  \hat{\vec{\delta}}_{\rm m}^{\rm t}\hat{\vec{\delta}}_{\rm l}(\vec{\theta},k,l)=
  \sum_{i=1}^{N_{\rm lp}}\sum_{j=k}^l[\vec{p}^{(i)}]^{\rm t}\vec{e}_j\delta_{\rm
    m}^{(i)}\delta_{\rm  l}^{(j)}\;,
\end{equation}
which on average is
\begin{equation}
  \Ave{\hat{\vec{\delta}}_{\rm m}^{\rm t}\hat{\vec{\delta}}_{\rm
      l}(\vec{\theta},k,l)}\approx
  \sum_{i=k}^lw_i\Ave{\delta_{\rm
      m}^{(i)}\delta_{\rm  l}^{(i)}} \;,
\end{equation}
where we define the weights $w_i\equiv[\vec{p}^{(i)}]^{\rm
  t}\vec{e}_i$. From the latter it becomes obvious that Eq.
\Ref{eq:product} can be made an estimator for the weighted l.o.s.
product of mass and light distribution within the selected redshift
range $[z_k,z_l]$, namely by the product map (PM) 
\begin{equation}
  \label{eq:productmap}
  {\rm PM}(\vec{\theta},k,l)\equiv
  \frac{\hat{\vec{\delta}}_{\rm m}^{\rm t}\hat{\vec{\delta}}_{\rm
      l}(\vec{\theta},k,l)}
  {\sum_{i=k}^lw_i}\;;
\end{equation}
which for statistically independent lens planes has the average
\begin{equation}
  \Ave{{\rm PM}(\vec{\theta},k,l)}=
  \frac{\sum_{i=k}^lw_i\Ave{\delta_{\rm
        m}^{(i)}\delta_{\rm  l}^{(i)}}}{\sum_{i=k}^lw_i}\;.
\end{equation}
We use product maps to obtain candidates for l.o.s. directions for
which features in the mass map could be related to stellar mass
(\mbox{${\rm PM}>0$}), directions that are significant in the mass map
but apparently uncorrelated to the stellar mass map (\mbox{${\rm
    PM}=0$}) or even anti-correlated (\mbox{${\rm PM}<0$}). The
candidates will be subject to further modelling of the radial density
profile to determine the probability distribution of mass along the
l.o.s., Sect.  \ref{sect:losmodel}.

\begin{figure}
  \begin{center}
    \psfig{file=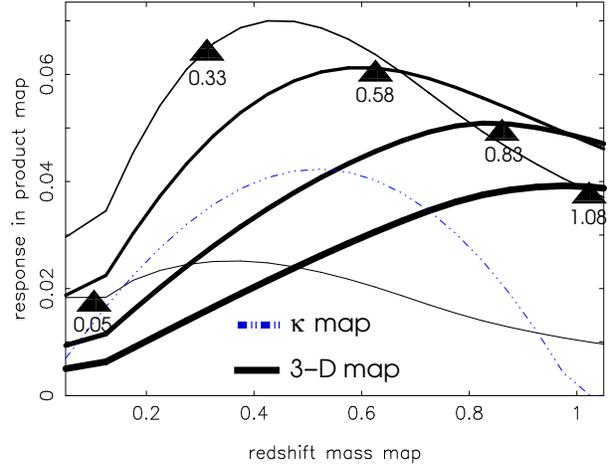,width=80mm,angle=0}
  \end{center}
  \caption{\label{fig:zconfusion}
    {A matter density fluctuation in the mass map still contributes to the product map even if it is at a different redshift ($x$-axis) as a fluctuation in the stellar mass map (redshift of filled triangle).  This is an effect of the broad radial smoothing kernel the Wiener map is subject to.  In an ideal world, the solid lines would be strongly peaked about the triangles, i.e. $[\vec{p}^{(i)}]^{\rm t}\vec{e}_j\approx\delta^{\rm K}_{ij}$. Thicker lines belong to higher stellar mass redshifts. For comparison, the dashed-dotted line shows the response we would expect for a purely 2D (convergence) map; the response curve is independent of the stellar mass redshift in this case.}}
\end{figure}

The product method has a limited radial resolution due to the broad
radial p.s.f. $\vec{p}^{(i)}$. For illustration, consider a l.o.s.
that has exactly one non-zero pixel $\Delta_{\rm l}$ in
$\vec{\delta}_{\rm l}$ on the $i$th lens plane. On the other hand, the
matter distribution may have the only non-zero pixel $\Delta_{\rm m}$
on the $j$th lens plane, hence at a different redshift. Despite an
obvious mismatch between light and mass, this case will contribute to
the product map with $\Delta_{\rm l}\Delta_{\rm m}[\vec{e}_i]^{\rm
  t}\vec{p}^{(j)}$ (Eq. \ref{eq:product}), if $\vec{p}^{(j)}$ is broad
enough to reach the $i$th lens plane.

In Fig. \ref{fig:zconfusion}, we show the response, or signal, in the
product map to a stellar mass point source at discrete redshifts $z=
0.05, 0.33,...1.08$ (shown with triangles) from a lensing mass spread
over the full redshift range \mbox{$0<z<1$}. Mass along the full
redshift range will contribute at some level to the product map, even
though it is not associated with the stellar mass point source. This
shows that our interpretation of the product map is non-trivial as
there is always considerable confusion present for the \mbox{STAGES}
map. However it does show that we have relatively little confusion
from low redshift lensing mass features to stellar mass redshift range
is \mbox{$z\gtrsim0.5$}. This is important as it means one can, in
principle, distinguish in the product map the supercluster structure
at $z\sim0.165$ from structures in the higher redshift
regime. {The response could be further improved by using less
  radial smoothing in the Wiener map by means of devising a smaller
  tuning parameter. This, however, also decreases the signal-to-noise
  in the product map.}

{A cross-correlation of a stellar map with a 2D convergence map,
  on the other hand, always produces the same response irrespective of
  the stellar mass redshift. This response is shown in
  Fig. \ref{fig:zconfusion} by the dotted-dashed blue (amplitude was
  rescaled for plotting purposes). Although the cross-correlation with
  the 3D Wiener map displays some improvement over a
  cross-correlation with a convergence map, it is unclear how the
  still broad response can be properly dealt with. For instance, a
  massive mass peak at low redshift could generate a strong response
  with a large stellar mass peak at high redshift.}

{To advance the 3D product map, it is conceivable to define a
  radial smoothing of the light map in order to optimise the response
  in the product map by substituting $\vec{e}_i$ by an appropriate
  vector. In addition, one could properly account for the $z$-bias so
  that the maximum response is at the location of the stellar mass
  peak or trough, which is not exactly the case here
  (Fig. \ref{fig:zconfusion}). We do not pursue such a strategy for
  this work, though, as we are not expecting a significant improvement
  for STAGES. In particular, we do not use a 3D product map to draw
  strong conclusions about radial matches of stellar mass and matter
  mass.  Instead, we solely employ the product map in combination with
  the Wiener S/N-map to identify interesting l.o.s. to be further
  analysed by a purely lensing method Sect \ref{sect:losmodel}.}

Note that smoothing the light map with the Wiener map p.s.f. before
multiplication, i.e. set $\vec{e}_i=\vec{p}^{(i)}$, results in a worse
response than the one adopted for this analysis; the response would be
the product of pairs of Wiener p.s.f. equalling their overlap,
i.e. $w_i=[\vec{p}^{(i)}]^{\rm t}\vec{p}^{(i)}$ in this case.

Finally, the signal-to-noise of the product map is assessed by
dividing Eq. \Ref{eq:productmap} by the variance seen in noise
realisations, which are obtained by randomising the positions of
galaxies in the $M_\ast$ map. Note that the signal-to-noise levels in
the product map can become considerably larger than in a pure lensing
map because for uncorrelated noise of the lensing and $M_\ast$ map the
signal-to-noise in the individual maps just multiplies.

\subsection{Executive Summary}
\label{sect:execsummary}

We have reviewed the theory presented in STH09, then presented details
of the extensions and new analysis tools. These tools were developed
for the first application of this method to real data.  Here we
highlight the new results presented above for different readers.

For readers optimising future survey designs, the most relevant
results are contained in Figs.~\ref{fig:zresolution} and
\ref{fig:minmass}, which show the best 3D mass resolution that can be
attained on the considered angular scales with space-based quality
data with a galaxy number density of $65$ galaxies per square arcmin
matched by $10$ galaxies per square arcmin with ground-based
photometric redshift estimates with a redshift error of
$\sigma_z/(1+z) = 0.02$.  Note these photometric redshifts are limited
at $z<1.4$ in the absence of near infra-red photometry to extend
reliably to higher redshifts.  This survey characterisation is similar
to planned deep imaging areas of space-based surveys with matching
photometric redshifts from shallower multi-colour optical surveys such
as PanSTARRS, DES or VST-KIDS.  These reveal rather relatively poor
expectations for attaining mass resolution in 3D using lensing alone
for anything other than the most massive structures (at least several
$10^{14} M_\odot h^{-1}$).  It therefore motivates future surveys to
invest in sufficiently deep photometric ground-based surveys such as
LSST and matching deep near infra-red photometry.
  
For readers focused on applying the STH09 method to a new data set,
Sect.~\ref{sect:radialpsf} and~\ref{sec:zbias} detail how to make the
non-trivial measure of $Z$-shift bias and the radial p.s.f from real
data that was not discussed in STH09.  Fig.~\ref{fig:zshiftbias} also
shows how the application of standard 2D mass mapping noise estimate
techniques to a 3D mass map biases the redshift measure with the
implication that a 3D S/N map, which has not previously been
calculated in the literature, is essentially a 2D map {for
  lensing surveys like STAGES. }  Section~\ref{sect:productmap}
therefore outlines a new product map method that enables one to
compare significant peaks detected in the essentially 2D S/N mass to
the 3D stellar mass distribution and the 3D lensing density
distribution. 

{We show that the product map offers theoretical improvements
  over a cross-correlation of a 3D stellar mass map with a 2D
  convergence map, but it remains unclear at this point how to clearly
  exploit this map. Nevertheless, we employ the product map to select
  interesting l.o.s. to be analysed further in the following.}

\section{Results}
\label{sec:results}

\begin{figure*}
 \begin{center}
   \psfig{file=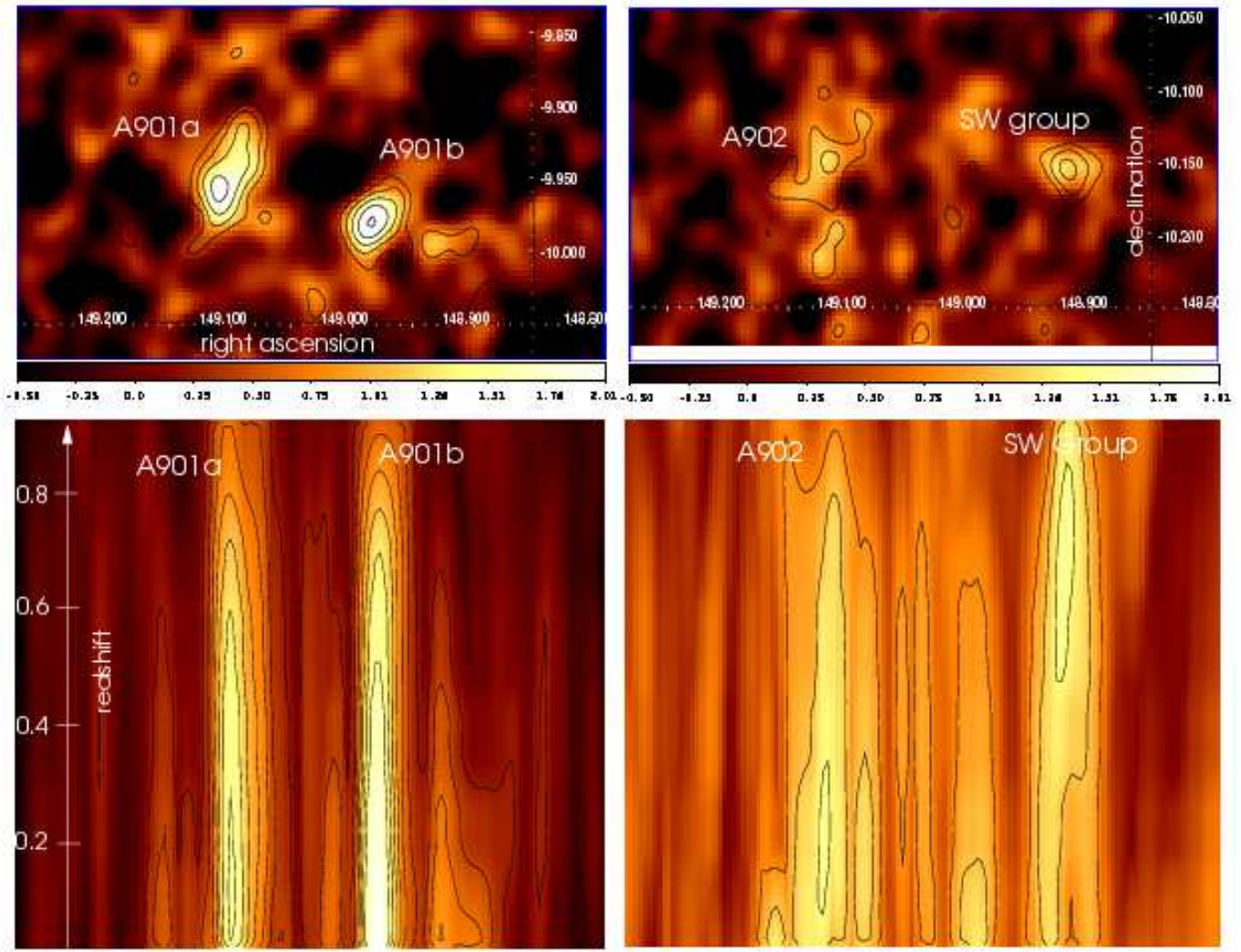,width=180mm,angle=0} 
   \caption{\label{fig:stages3dmap}
     {Different projections of the Wiener mass map (\emph{top row}:
       projection onto the sky; \emph{bottom row}: radial projection
       of the area in the top row; $z$-shift bias corrected) The
       colour scales shows the maximum density contrast in the map along
       the projection axis. The contours in the top panels correspond
       the S/N levels $2.5\sigma$, $3\sigma$, $4\sigma$, $5\sigma$ and
       $6\sigma$ as determined from noise realisations of the
       data. Contours in the bottom panels are density contrast
       levels included to highlight the structure in the radial
       map. Structures appear radially stretched due to the impact of
       the Wiener filter smoothing kernel.}}
 \end{center}
\end{figure*}

In this section, we present the results of our 3D reconstruction of
the STAGES A901/902 supercluster field and our analysis of that
reconstruction. We start by presenting the raw Wiener reconstruction
in Fig. \ref{fig:stages3dmap}, splitting the field into two parts: the
northern half of the field, comprising A901a/b (left column), and the
southern part, comprising A902 and the SW group (right column). In the
upper panels we present the 3D density map projected on the plane of
the sky showing with high significance the four structures in the
supercluster. This projection shows the same result as the 2D
reconstruction in \cite{Heymans08} and we overlay signal-to-noise
contours obtained by comparing the mass map to noise realisations.
Dividing by the noise in the 3D mass map, changes the $z$-shift bias
in the new resulting map, rendering it essentially a 2D map for
\mbox{STAGES}. For this reason, the S/N map is shown only in 2D as
maximum along a l.o.s. direction. In the lower panels, we present the
3D map now projected along the declination. This reconstruction shows
the redshift-density distribution convolved with the the radial p.s.f.
of the map, shown in Fig.~\ref{fig:radpsf}, corrected for $z$-shift
bias, shown in Fig.~\ref{fig:zshiftbias}. As this radial p.s.f is so
broad, structures are radially spread out in the map and overlap in
redshift. Nevertheless, in cases where structures do not lie in close
projection, once the $z$-shift bias is corrected local maxima are
unbiased estimators of the redshift of mass concentrations. Whilst
unbiased however they have limited redshift accuracy, reflected by
Fig. \ref{fig:zresolution}.

\begin{figure*}
\begin{center}
   \begin{tabular}{cc}
     \psfig{file=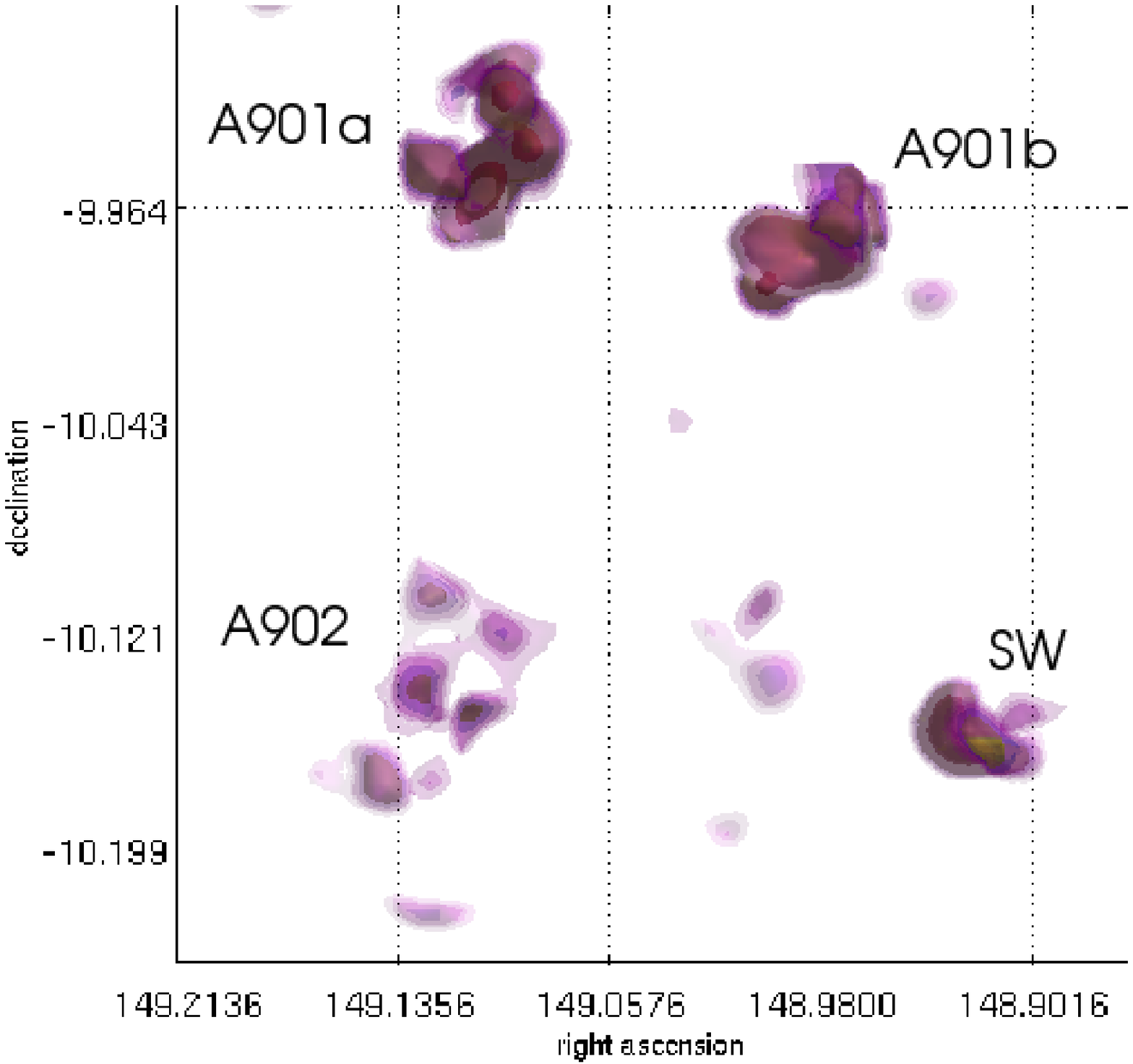,width=90mm,angle=0} &
     \parbox{60mm}{\vspace{-5cm}\caption{\label{fig:pretty}}
       3D Dark Matter Visualisation. Representation of
       the local l.o.s. maxima in the raw \mbox{3D} Wiener map
       of STAGES, obtained from
       Fig. \ref{fig:stages3dmap}.  
       This visualisation of the 3D density field is an noisy estimator of the
       true distribution of mass concentration. The radial spread of
       structure does not reflect the redshift uncertainty.}
     \\
     \multicolumn{2}{c}{\psfig{file=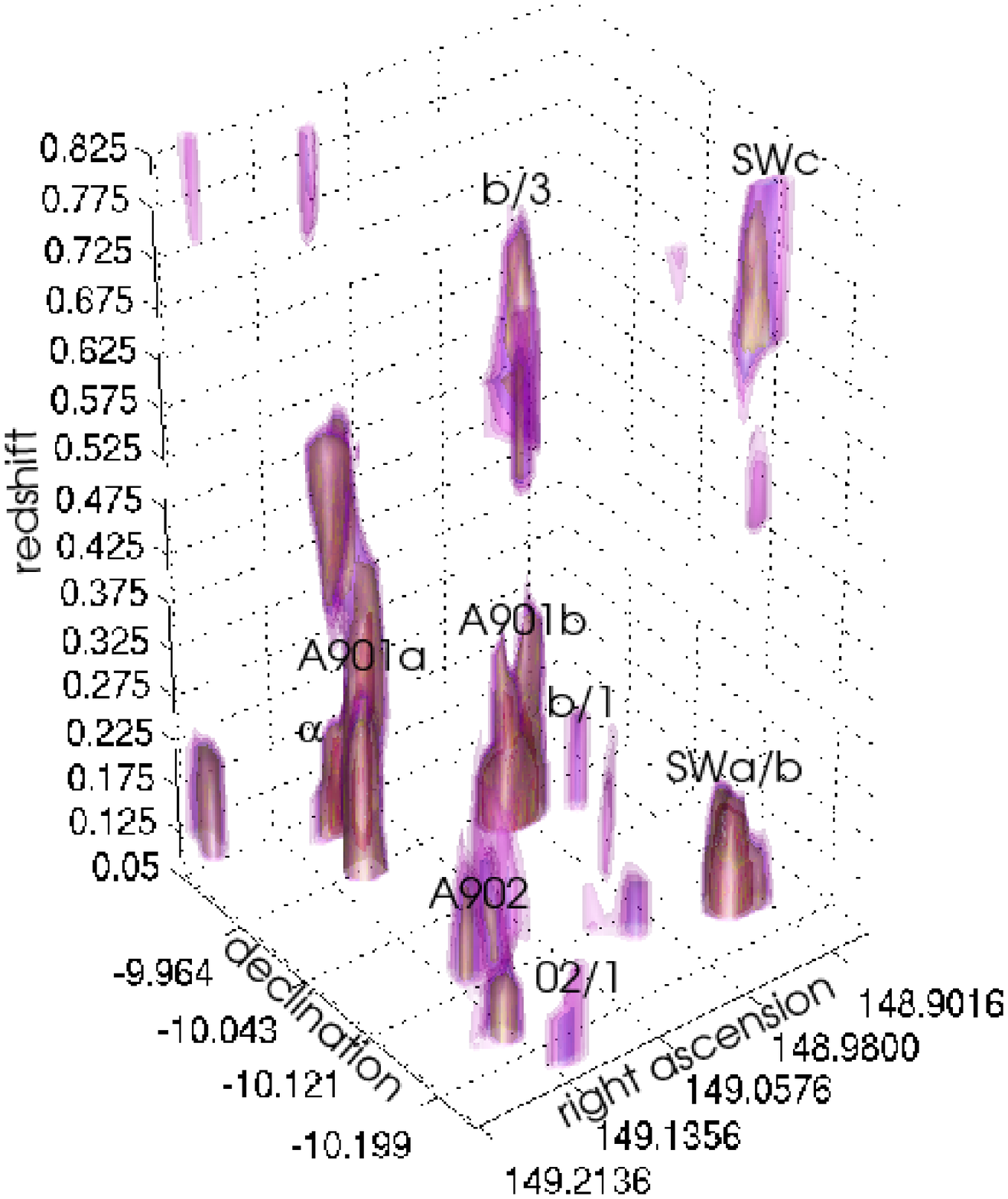,width=110mm,angle=0}}
   \end{tabular}
 \end{center}
\end{figure*}

{In order to better visualise the 3D matter distribution from
  the raw Wiener map, we make the general assumption that the local
  maxima are good estimators of the peaks in the density field.  We
  then create a 3D map retaining only the density pixels that are a
  local maxima along each l.o.s..  We then smooth the visualisation
  with a Gaussian kernel $(0.2\,\rm arcmin)^2$ in r.m.s. size on the
  sky and $0.025$ in redshift with the result shown in
  Fig.~\ref{fig:pretty}.  This cleaning process produces a striking
  image of the 3D density field revealing structure at high redshift
  (marked SWc and b/3) behind the foreground supercluster (marked
  A901a, A901b, A902, SWa/b).  It is this realisation of the noisy 3D
  density field that should be compared with other 3D mass
  reconstructions in the literature \citep{MasseyNat,
    Taylor04}. {A one-to-one comparison is still not easy,
    though, since other reconstructions show the estimated
    gravitational potential.}  We strongly caution the reader over the
  direct use of 3D lensing maps, however, as they are noisy
  estimators of the redshift distribution and contain no error
  information.  In particular, owing to the broad smoothing kernels,
  structure along the same l.o.s. can merge in this visualisation to
  produce a density profile with a local maximum at an intermediate
  redshift. As our detailed analysis of the raw Wiener map will show
  however, after the radial p.s.f has been correctly accounted for,
  the higher redshift structures picked out in this visualisation are
  still seen with a high probability.}

\subsection{3D comparison of mass and light}
\label{sect:comparison}

One of the main science drivers for 3D mass mapping is to investigate
how well the total mass traces light or stellar mass in 3D. We use
the product map technique outlined in Sect. \ref{sect:productmap} to
compare total mass and stellar mass in redshift slices and identify
interesting l.o.s. in our 3D map.

We use $17$- band \mbox{COMBO-17} stellar mass $M_\ast$ estimates for
the full galaxy sample down to \mbox{$R<24\,\rm mag$}. The
$M_\ast$-map is produced by summing the stellar masses $M_i$ of
galaxies within the same angular bin and lens plane (pixel).  Bins
containing no galaxies have a stellar mass of zero. This matrix of
binned masses is divided by the average pixel mass,
$\overline{M}=\sum_i M_i/N_{\rm p}$ where $N_{\rm p}$ is the total
number of unmasked pixels. All pixels of the same l.o.s. are flagged
unmasked if there is at least one pixel along the l.o.s. containing
one or more galaxies. The majority of masked pixels are found at the
edges of the field. The effect of masking in the central part of the
map is hence small. The normalised $M_\ast$-map of the lens plane is
in the unmasked regions, up to a constant, an estimator for the
$M_\ast$ density contrast, $M(\vec{\theta})/\overline{M}-1$. In a next
step, the average $M(\vec{\theta})/\overline{M}$ is subtracted from
all pixels.

\begin{figure*}
 \begin{center}
   \psfig{file=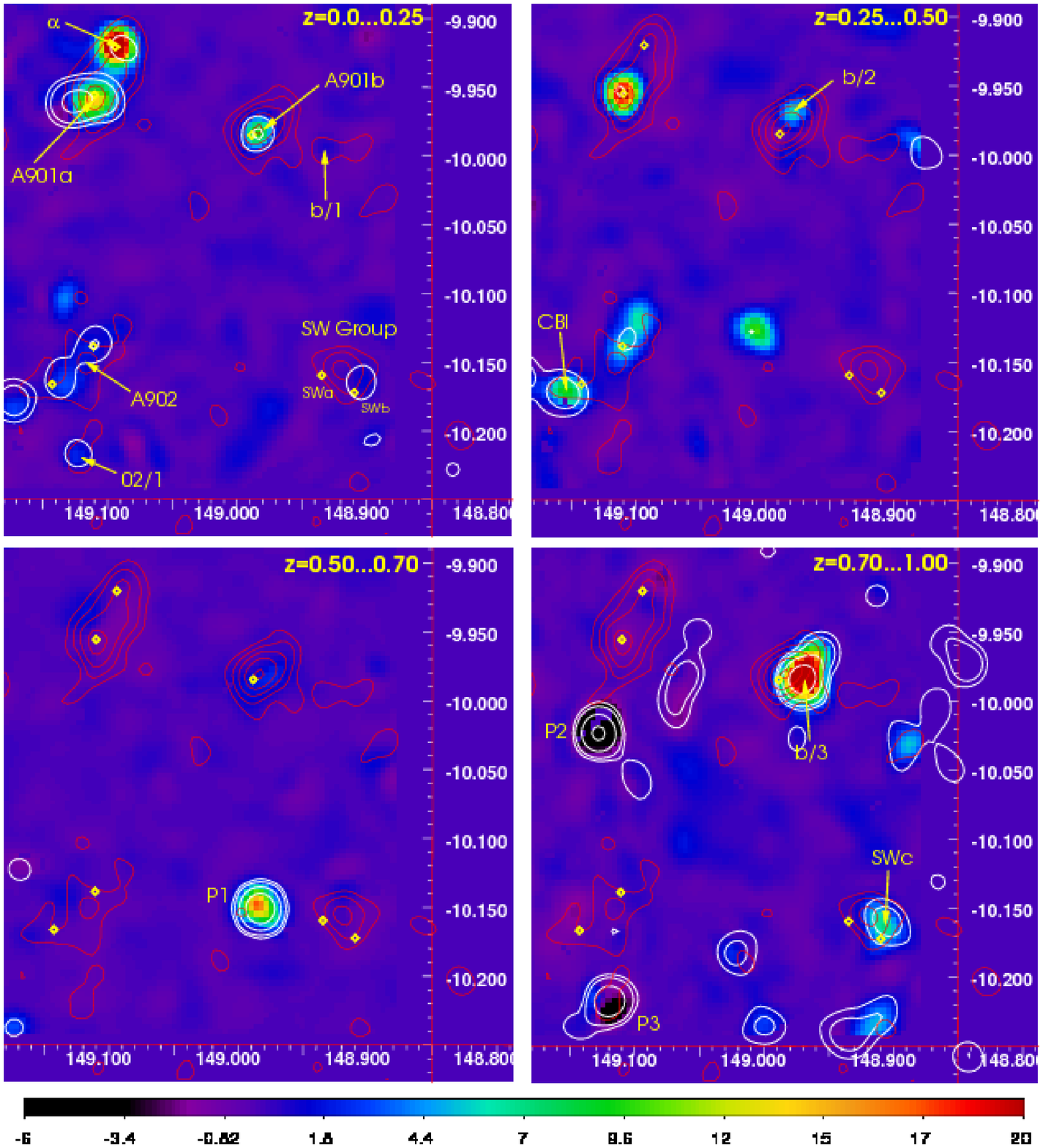,width=175mm,angle=0}
 \end{center}
 \caption{\label{fig:zoomin}
   {Total-Stellar Mass Product maps obtained by multiplying a stellar mass map with the Wiener mass reconstruction within four redshift slices (top left for $z$-range). {Intensity levels in the background show the signal-to-noise in the product map with black regions having the most negative values and red the most positive levels.}  Signal-to-noise mass density contours are shown in red with levels $2.5\sigma$, $3\sigma$, $4\sigma$, $5\sigma$ and $6\sigma$, depicting the maximum significance along a l.o.s. in the lensing mass map. Note the source of the signal may be at any redshift as the S/N maps are almost completely spread out in redshift. White contours highlight regions of high stellar mass $M_\ast=10^{10},2\times10^{10},5\times10^{10},10^{11} \msol ,h^{-1}$ in the $M_\star$ map smoothed with a 1.2 arcmin Gaussian kernel. The map pixel volume increases towards greater redshift. Open diamonds are reference points marking prominent bright cluster galaxies in A901a/b, A902 and the SW group. Numbers on the $x$- and $y$-axis denote the right ascension and declination, respectively (J2000.0). }}
\end{figure*}

In Fig. \ref{fig:zoomin} we present the Total-Stellar Mass Product Map
of the STAGES field within four redshift slices $z\in[0,0.25]$,
$z\in[0.25,0.50]$, $z\in[0.50,0.70]$ and $z\in[0.70,1.00]$, where the
redshift bin width is significantly larger than the photometric
redshift error $\sigma_z = 0.037$. The colour-scale shows the product
map level normalised by the variance in noise realisations; black
through purple regions have negative values and blue through red
have positive values. We can see the majority of the density map is
positively correlated with the stellar mass map. Overlaid in red are
S/N contours of the density map. As shown in Sect.~\ref{sec:zbias},
the S/N maps are almost completely spread out in redshift and so we
use the same S/N map in all redshift slices to identify regions in the
plane of the sky where we find a significant detection of mass. The
3D stellar mass distribution is shown with white contours. We find a
clear correlation of mass and stellar mass in the $z=0.165$ A901/902
supercluster in the lowest redshift panel (upper left) and also the
CBI cluster at $z=0.46$ (upper right).

The interpretation of this map is however non-trivial as the radial
p.s.f. of the map causes overdensities in the mass map to contribute
to the product map even when they are at a different redshift to the
concentration in the stellar mass map
(Sect. \ref{sect:productmap}). We can therefore only use this product
map to identify interesting l.o.s., which we then follow up with a
detailed analysis that incorporates knowledge of the radial p.s.f. of
the Wiener map. We identify interesting l.o.s. as follows. First, we
only consider pixels in the \emph{total mass map} that are detected
with a \mbox{${\rm S/N}>3$} in projection. We then choose all pixels
where we find a peak positive {or negative} correlation of
significance greater than $1\sigma$ with the stellar mass. {The
  majority of the lines of sight above the $3\sigma$ threshold exhibit
  a positive correlation in the product map. According to the noise
  peak statistics in Appendix \ref{sect:noisepeaks}, however, we
  expect at least some of the matches below \mbox{$\lesssim3.5\sigma$}
  to be confusions of random lensing map fluctuations with the 3D
  stellar mass map.}
{We find that the cases of negative correlation in the lensing
  map with \mbox{${\rm S/N}>3$} are all behind foreground structures
  A901a, A901b, A902 and SW, which are likely to be produced by the
  leakage effect, outlined in Sect. \ref{sect:productmap}.  With the
  exception of A901a, these negative correlations have a product map
  signal-to-noise of \mbox{$<3\sigma$} and are therefore statistically
  insignificant.  They are all included in the analysis of the
  selected l.o.s. with positive correlations in the foreground, and
  therefore automatically included in the following analysis.}  Our
ten chosen l.o.s. are listed in Table \ref{table:peakpos} with
identifiers based on the known foreground $z=0.165$ structures in this
group, A901a, A901b, A902, SW, and also the higher redshift CBI
cluster at $z=0.46$.\footnote{Note that SWa/b, galaxies belonging
    to a known group at $z=0.165$, did not make it into the table
    because the lensing significance in direction of PM peaks
    corresponding to SWa/b is too low; for SWb it is
    $2.5\sigma$. However, SWb lies within the range of the SWc pencil
    beam analysed later on.}

\begin{table}
  \caption{\label{table:peakpos} 
    Compiled catalogue of l.o.s. candidates of positive mass-light
    correlations with {projected l.o.s. significance higher than
      $3\sigma$ in the lensing map (shown as S/N$_\kappa$); 
      S/N$_{\rm PM}$ lists the corresponding signal-to-noise in the
      product map. Lines of sight with no product map match are
      not listed.} The equatorial coordinates (J2000.0) are the peak positions in
    the product map, redshift constraints are estimates with $68\%$ credibility
    from a thorough l.o.s. modelling in Sect. \ref{sect:losmodel}. A
    dagger$^\dagger$ denotes redshift estimates from a two component
    model rather than a one-component mass model.}
  \begin{center}
    \begin{tabular}{@{}lccc@{}cc}
      \hline\\
      Peak & REC & DEC & S/N$_\kappa$~~ & S/N$_{\rm PM}$ & $z$\\
             &  [hh:mm:ss] & [$^\circ:^\prime:^\pprime$]\\
      \hline\hline\\
      A901b & 09:55:56.1 & --09:59:06.5 & $6.5$ & $9.0$ & $0.12_{-0.08}^{+0.10}$\\
      A901a$^\dagger$ & 09:56:25.3 & --09:57:27.8 & $5.8$ & $14.5$ & $0.17_{-0.12}^{+0.15}$\\
      A901b/3$^\dagger$ & 09:55:53.9 & --09:58:54.6 & $5.1$ & $32.8$ &
      $0.57_{-0.30}^{+0.35}$ \\
      A901b/2 & 09:55:54.4 & --09:58:03.1 & $4.8$ & $6.2$ & $0.28_{-0.15}^{+0.20}$\\
      A901$\alpha^\dagger$ &  09:56:22.3 & --09:55:09.1 & $4.7$ & $31.4$ & $0.42_{-0.30}^{+0.40}$ \\
      SWc$^\dagger$ & 09:55:39.9 & --10:08:55.7 & $4.1$ & $7.0$ & $0.72_{-0.30}^{+0.25}$\\
      A902$^\dagger$ & 09:56:28.1 & --10:09:01.1 & $3.9$ & $1.9$ & $0.28_{-0.15}^{+0.25}$\\

      A901b/1$^\dagger$ &  09:55:44.3 & --09:59:26.1 & $3.2$ & $1.0$ &$0.17_{-0.05}^{+0.30}$\\
      CBI  & 09:56:36.5  & --10:10:25.7 & $3.1$ & $9.6$ & $0.68_{-0.45}^{+0.30}$\\
      A902/1$\dagger$ & 09:56:29.0 & --10:13:02.7 & $3.1$ & $2.2$ &$0.17_{-0.12}^{+0.25}$
    \end{tabular}
  \end{center}
\end{table}

Before continuing with our detailed line of sight modelling of the ten
structures marked in the product map, we briefly discuss other strong
features.  In the two highest redshift slices, we find two stellar
mass overdensities at $M_\ast = 10^{11}M_\odot$ (with 1.2 arcmin
smoothing). In the $z\in[0.50,0.70]$ bin, we see a strong positive
correlation P1 of total mass and stellar (shown white) where the total
mass is detected at $2.5\sigma$. In the $z$-bin $z\in[0.70,1.00]$, we
find significant anti-correlation P2 ($\sim6\sigma$ in the product
map), which also is a stellar mass overdensity matched by a lensing
mass underdensity. The $M_\ast$ overdensity region is found to be
overlapping with gaps produced by three stars in this area, which may
be interfering with the lensing map. Moreover, as shown in
Fig.~\ref{fig:minmass}, at these high redshifts we would only expect
to reliably detect haloes of mass $M_{200}\sim2\times10^{14} M_\odot$,
a higher mass range than what we would expect for these stellar mass
overdensities. The same argument holds for the feature P3. These
stellar mass overdensities therefore demonstrate the effect of noise
in our mass maps where we are detecting mass with lensing only at the
$2.5\sigma$ level.  We also note that we find no significant
($\gtrsim3.5\sigma$) lensing mass overdensity matched by a stellar
mass underdensity in the entire 3D map (``dark clump'').

\subsection{Line of sight modelling}
\label{sect:losmodel}

Using the product maps in Fig.~\ref{fig:zoomin}, we identify
interesting l.o.s. in the map to perform a more detailed analysis of
the 3D structure that takes fully into account the radial p.s.f. of
the Wiener map. For each l.o.s., listed in Table~\ref{table:peakpos},
we model the matter distribution by assuming that it consists of one
individual non-zero mass halo \mbox{$\propto1+\delta_0$} on the lens
planes.  The model profile in the convolved 3D Wiener map is given by
\begin{equation}
  \label{eq:onepixelmodel}
  \hat{\vec{\delta}}_{\rm m}(i,\delta_0)=
  \delta_0\vec{p}^{(i)}\;,
\end{equation}
where \mbox{$\delta_0>-1$} is the density contrast of the non-zero
pixel on the $i$th lens plane; all other pixels have $\delta_{\rm
  m}^{(i)}=0$ which amounts to average density in the corresponding
epoche. The radial matter density $\hat{\vec{\delta}}_{\rm m}$ in
direction $\vec{\theta}$ of the Wiener map with noise covariance
$\mat{N}$ enables us to write down the posterior likelihood of the
mass pixel lens plane index $i$ and its contrast
$\delta_0$: \begin{eqnarray}
  \label{eq:onedensity}
  &&  \!\!\!\!\!\!\!\!\!P(i,\delta_0|\hat{\vec{\delta}}_{\rm m})
  \propto   H(\delta_0+1)\times\\\nonumber
  && \exp{\left(-\frac{1}{2}
      \left[\hat{\vec{\delta}}_{\rm m}-\hat{\vec{\delta}}_{\rm
          m}(i,\delta_0)\right]^{\rm t}\mat{N}^{-1}
      \left[\hat{\vec{\delta}}_{\rm m}-\hat{\vec{\delta}}_{\rm
          m}(i,\delta_0)\right]\right)}\;,
\end{eqnarray}
where the Heaviside function prior $H(x)$ asserts a positive density
$1+\delta_0\ge0$.

\begin{figure*}
 \begin{center}
   \psfig{file=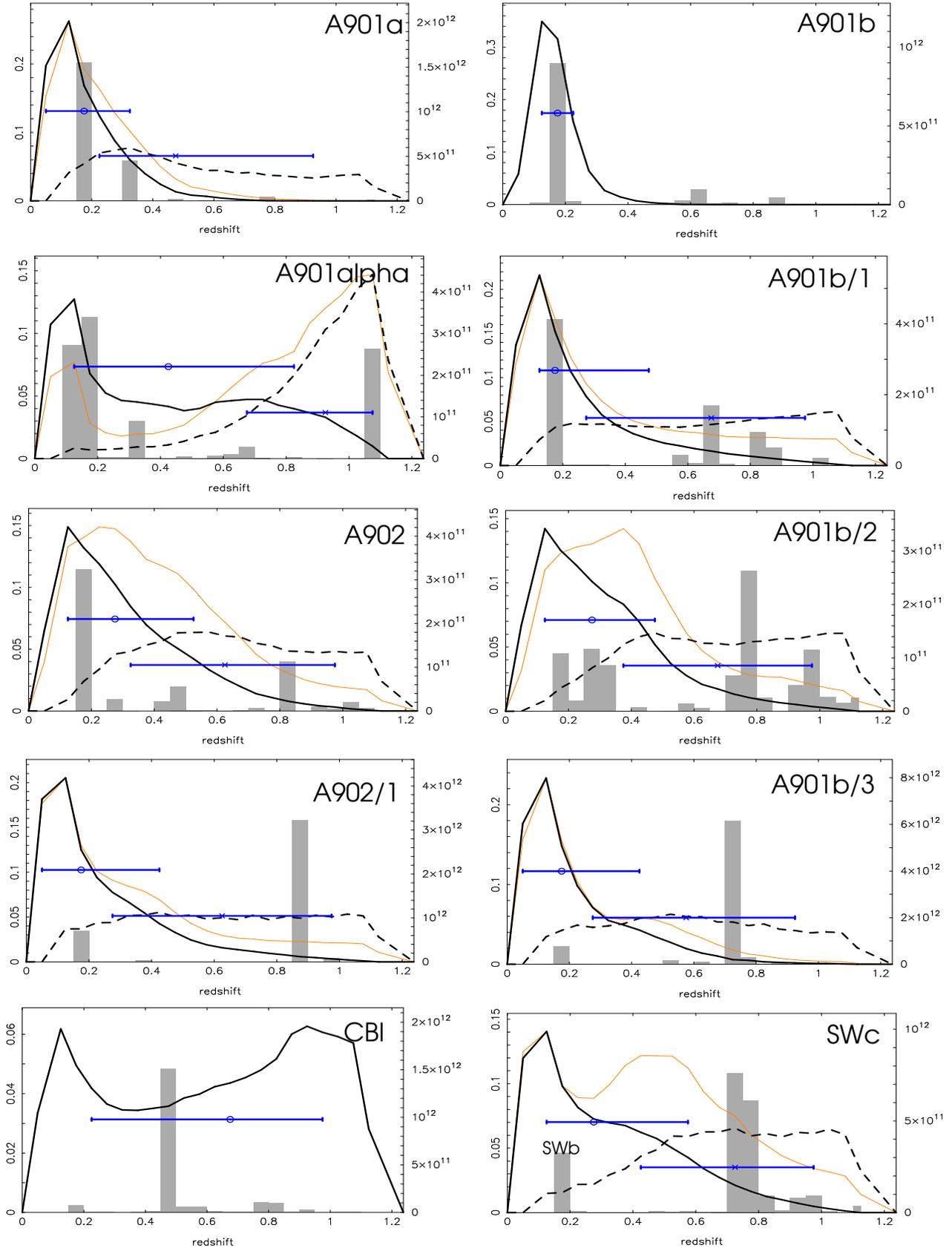,width=168mm,angle=0}
 \end{center}
 \caption{\label{fig:los} Probability distribution of redshifts in a
   one-component (only one black solid line or orange solid line) or a
   two-component (solid and dashed black line) mass model for the
   radial density profile into different directions of
   \mbox{STAGES}. Grey bars indicate the stellar mass $M_\ast$ within
   a pencil beam of $\sim40\,\rm arcsec$ radius in units of $\msol
   h^{-1}$ (numbers on right edge). {The blue error bars denote a
   $68\%$ confidence regions about the median redshift of the mass 
   concentration.}}
\end{figure*}

Marginalising over $\delta_0$, we employ the MCMC technique to
construct a probability distribution $P(i|\hat{\vec{\delta}}_{\rm
  m})$ for each lens plane $i$, as shown as black lines in Fig.
\ref{fig:los}. This probability distribution gives the true
probability of a density peak positioned along each l.o.s. at a given
redshift. This result correctly accounts for the broad radial p.s.f
that has hampered the interpretation of the maps presented up to this
point. The blue error bars denote a $68\%$ confidence regions about
the median redshift of the non-zero pixel. For comparison, the density
contrast from the $M_\ast$ map for the same pencil beam is shown as
grey filled bars; the beams have a radius of $\sim40\,\rm arcsec$. The
bars are normalised such that the maximum density reaches the half
height of the diagram.

\begin{figure}
  \begin{center}
    \psfig{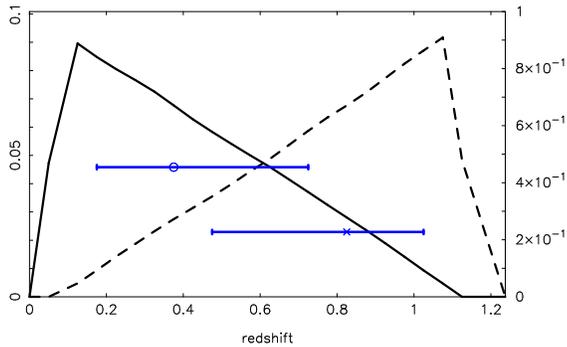}
  \end{center}
  \caption{\label{fig:losnull}
    {
      Probability distribution for redshifts in a two-component
      model in the absence of any signal (null distribution). The
      distributions for $z_1$ and $z_2$ are not top-hats due to the
      a-priori condition $z_1<z_2$.}}
\end{figure}

Motivated by the results of the product map in Fig.~\ref{fig:zoomin},
we also test an alternative two-halo mass model for most of the l.o.s.
, which seem to require a more complex modelling when comparing the
one-component p.d.f. to the $M_\ast$ distribution in the pencil
beam. The posterior likelihood of the two-component model is similar
to Eq. \Ref{eq:onedensity}, but now contains two positions $i_1$ and
$i_2$ and two densities. As additional prior, we demand that
$i_1<i_2$. The resulting marginalised probability distributions are
shown as black solid and dashed lines, the two error bars denote the
constraints on redshifts in the two-component model. The orange solid
line shows the probability distribution for the one-component model.
Note that in a two-component model fit with no signal in the data, the
probability densities of both components are not simply flat: Due to
the adopted prior $z_1<z_2$, the null signal distribution has a
saw-tooth shape as in Fig. \ref{fig:losnull}, which is produced due to
the correlation of the probability distributions of $z_1$ and $z_2$.

In Fig.~\ref{fig:los} we see good agreement between the peaks in the
density redshift probability distribution and the overdensities in
stellar mass. We also see clear evidence for l.o.s. projections in the
stellar mass distribution which have a significant probability in the
3D density map. In almost all cases, a two-component model is not
clearly preferred over a one-component model with the slight exception
of A901$\alpha$ that prefers a two-component model (d.o.f.=$18$) with
$\Delta\chi^2=1.90$ ($\sim68\%$ confidence) as compared to a
one-component model (d.o.f.=$20$).  The best-fit one-component model
of \mbox{SWc} has \mbox{$\chi^2=17.23$} as opposed to
\mbox{$\chi^2=16.70$} for the two-component model, which is
statistically insignificant.  For \mbox{SWc}, however, a model with
all mass concentrated at $z=0.165$, the distance of \mbox{SWa/b}, is
excluded with $\sim68\%$ confidence ($\chi^2=18.69$; d.o.f.=$21$).
Combining this information with our knowledge of the galaxy
distribution, we favour the two component model which reveals a new
high redshift structure SWc at $z\sim0.7$, lying in projection with
the foreground SW group at $z = 0.165$.

\section{Discussion and Conclusions}
\label{sec:discuss}

This paper uses 3D lensing data from the STAGES project to study the
spatial distribution of matter density and galaxies in the Abell 901/2
supercluster field. We have presented a non-parametric lensing-based
3D mass density reconstruction. This application of the 3D mapping
algorithm to data is the first of its kind as earlier attempts
\citep{MasseyNat, Taylor04} to do cosmography with 3D lensing data
focused on a reconstruction of the gravitational potential rather than
the mass density itself. In addition to directly mapping the 3D
density field, the new algorithm presented in this paper accounts for
redshift bias, a term previously neglected in the literature, and it
optimally includes all lensed sources including those without reliable
photometric redshift estimates.\footnote{{In principle, the
    $z$-shift bias is also be present in previous Wiener filter maps
    of the 3D gravitational potential. That it was not recognised in
    the method of \citet{Taylor04} may be due to the difference in the
    Wiener filter, which processes all lines of sight individually,
    ignoring correlations of statistical errors between distinct lines
    of sight. This is suboptimal in terms of signal-to-noise in the
    resulting map, but on the other hand might result in a less severe
    $z$-shift bias.}}

We present both our raw Wiener reconstruction of the STAGES 3D
density field in Fig.~\ref{fig:pretty} and a cleaned visualisation of
the 3D field in Fig.~\ref{fig:stages3dmap}. Using the visualisation
is a valid tool to describe the structure in the field, in particular
because, in this instance, the $z$-shift bias has been correctly
removed. Further it is this cleaned map which should be compared to
previous 3D gravitational potential maps in the literature where the
displayed isophotal potential levels have been optimised to highlight
structures. However we urge caution in using any of these cleaned
visualisations for scientific analysis as they are very noisy
estimators of the density field and do not show the redshift error in
the map or account for the broad radial smoothing of the field, shown
in Fig. \ref{fig:radpsf}. However we have created a striking
visualisation tool of the 3D dark matter Universe whose impact should
not be underestimated even if its direct use for scientific analyses
is {somewhat} limited. Note, as discussed in STH09, for deeper
observations and higher number densities of sources, the radial
smearing effect is reduced but it will always be present as a generic
feature of this Wiener filter 3D lensing method, no matter how good a
data set is analysed. {A future improvement, albeit not explored
  for this paper, may be to regularise the Wiener filter, e.g. by a
  prior that forces structure to be confined to one lens plane.}
{Although this resulted in a more realistic estimate for the 3D
  matter distribution, it would not change the considerable
  uncertainties in the visualisation.}

From the raw 3D Wiener lensing reconstruction we infer that most of
the structure in the STAGES field belongs to the Abell supercluster at
\mbox{$z\sim0.165$} since most l.o.s. density profiles in the map are
locally peaked at around this redshift. There are a few exceptions
however, of which the most prominent is in the SW group region. Here,
we find a strong peak at around \mbox{$z\sim0.7$} apparent in the raw
Wiener map. 
  
In order to interpret the Wiener map result we have compared the total
mass to the stellar mass in the field using a product map technique
outlined in Sect.~\ref{sect:productmap}. In the resulting product map
in Fig. \ref{fig:zoomin}, significant matches or mismatches of density
fluctuations between total mass and $M_\ast$ from the selected
redshift range are identified and followed up with a more detailed
analysis in Sect.~\ref{sect:losmodel}. The conclusions we can draw
from this analysis are as follows:

\begin{enumerate}

\item {We find no detection of ``dark clumps'' in the 3D map,
    i.e. matches of lensing mass overdensities with stellar mass
    underdensities, which have been discussed in the literature
    \citep{2006A&A...454...37V, 2000A&A...355...23E, MasseyNat}.}

\item As expected from Fig. \ref{fig:zresolution}, A901b is well
  confined in redshift by the 3D lensing data, and the estimated
  matter distribution in the pencil beam fits well to the $M_\ast$
  distribution.

\item A901a, the second massive cluster in the field is less well
  constrained in redshift, possibly explained by the presence of an
  extra mass peak at around $z\sim0.3$ as seen in $M_\ast$ and also
  highlighted in the product map (Fig. \ref{fig:zoomin}, top right
  panel).

\item The SW group pencil beam is equally well fit by a one or
  two-halo model in projection. However the single-halo model exhibits
  an additional peak at higher redshift and is slightly inconsistent
  with all mass concentrated at $z=0.165$. We therefore conclude that
  along this l.o.s. there are indeed two structures; the well
  documented foreground SW group at $z=0.165$ and a second possible
  mass concentration at $z=0.7\pm0.3$, which is confirmed by a
  corresponding stellar mass peak. {SWc is also the most
    prominent high-$z$ peak in the raw Wiener map. This finding was
    missed in the past 2D lensing analyses of
    \citet{Gray02,Heymans08} and the smoother 3D map of the
    gravitational potential in \citet{Taylor04}.}
  
\item We find a similar result for the the A901$\alpha$ X-ray group
  known to be infalling on A901a at a redshift of $z\sim0.17$. In this
  case the one-halo model again rules out a low-redshift structure at
  moderate significance, suggesting that the equally well fit two-halo
  model best represents the data with widely scattered matter behind
  $z=0.165$ as indicated by the $M_\ast$ map.

\item A902 is equally well fit by a low-redshift one-halo model
  consistent with the known low-redshift of the cluster, or a two-halo
  projection model with one at $z\sim0.2$ and one at $z\sim0.8$.

\item CBI, clearly visible as $M_\ast$ peak at $z=0.46$, is weakly
  detected in the mass map, which indicates a mass feature at a
  consistent redshift $z=0.68_{-0.45}^{+0.30}$.

\item We also find evidence for mass concentrations behind the A901b
  cluster termed as A901b/2 and A901b/3. The latter is more prominent
  in the product map, concentrated at $z\sim0.7$.

\end{enumerate}

{Note that none of these conclusions could be drawn from the
  purely 2D analysis of \citet{Heymans08}. It is the non-parametric,
  purely lensing-based modelling of selected l.o.s.,
  Fig. \ref{fig:los}, that has to be used to investigate these
  significant features. In contrast to our 3D Wiener map and
  comparable visualisations
  \citep{hukeeton02,bacontay,Taylor04,MasseyNat,2011ApJ...727..118V},
  information on the redshift uncertainty of lensing mass peaks are
  directly visible in this Bayesian approach. From here we can see
  that even for a survey as good as STAGES radial distances of objects
  at higher redshifts are usually weakly constrained, most strikingly
  visible for the background objects in a two-component fit. This
  confirms the theoretical estimate in Fig. \ref{fig:zresolution},
  which is based on SIS matter haloes. Nevertheless, the influence of
  background matter on the A901b, A902 and especially SW group lensing
  signal is visible in Fig. \ref{fig:los}; compare the foreground
  p.d.f. in the two-component model to the p.d.f. of the one-component
  model. This influence is hidden in a 2D lensing analysis. }

\begin{table}
  \caption{\label{tab:nfwfit} {Constraints on the NFW virial masses
      $M_{200}$ and the virial radii  $r_{200}$ for A901b and SWa,b in a
      simple model fit to the 3D lensing data and additionally
      A901b/2-3 and SWc in an updated model with more components. The
      redshift of the foreground mass peaks is fixed with
      $z=0.165$ and with $z=0.8,0.75,0.75$ for A901b/2-3 and
      SWc, respectively. Within the fit an uncertainty of $\Delta
      z=0.05$ is factored in. Quoted are $68\%$ credibility regions
      about the mean of the posterior likelihood. As fiducial
      cosmology here $\Omega_{\rm m}=0.3$ and $\Omega_\Lambda=0.7$
      were assumed.}}
  \begin{center}
    \begin{tabular}{lcccc}
      & \multicolumn{2}{c}{Simple model} 
      & \multicolumn{2}{c}{Updated model}\\      
      Peak &
      $r_{200}$ & $M_{200}$ &
      $r_{200}$ & $M_{200}$\\
      & $[h^{-1}\rm kpc]$ & $[h^{-1}\rm 10^{13}M_\odot]$ 
      & $[h^{-1}\rm kpc]$ & $[h^{-1}\rm 10^{13}M_\odot]$\\ 
      \hline\\
      A901b   & $1140_{-122}^{+110}$  & $16.9_{-4.6}^{+5.4}$  & $1086_{-133}^{+112}$ & $14.6_{-4.4}^{+5.0}$\\
      A901b/2 & -- & -- & $376_{-163}^{+157}$ & $3.3_{-2.3}^{+4.8}$ \\
      A901b/3 & -- & -- & $325_{-161}^{+182}$  & $2.4_{-1.8}^{+5.1}$ \\
      \\
      SWa   & $507_{-228}^{+211}$ & $2.2_{-1.6}^{+3.1}$ & $481_{-221}^{+214}$ & $2.0_{-1.4}^{+2.9}$ \\
      SWb   & $496_{-215}^{+204}$  & $2.1_{-1.4}^{+3.0}$ & $486_{-210}^{+198}$ & $1.9_{-1.3}^{+2.7}$ \\
      SWc   & -- & -- & $290_{-155}^{+172}$ & $1.8_{-1.4}^{+3.8}$
    \end{tabular}
  \end{center}
\end{table}
 
All in all the modelling {and its comparison to the stellar mass
  distribution} confirms that the cleaned visualisation of the 3D
lensing map correctly identifies {previously known structure}
\citep[e.g.][]{Heymans08} at low redshift, but also seems to identify
hitherto {\emph{unknown structure}} at larger redshift, most
notably A901b/3 and SWc.  These overlapping mass concentrations are a
possible source of error when it comes to mass determinations by
lensing, a method that naturally measures the integrated mass within
cylinders along the l.o.s. {To gauge this effect, we fit
  Navarrow-Frenk-White (NFW) matter density halo profiles
  \citep{1997ApJ...490..493N,2000ApJ...534...34W} with fixed redshifts
  and positions on the sky to the 3D lensing data. A method similar
  to the one in \citet{Heymans08} is employed and hence not detailed
  here; the redshift p.d.f. of our shear catalogue subsamples are
  utilised, and constrains from all subsamples are combined. In one
  fit assuming a simpler mass model, we include SWa/b, A901b and
  A901$\alpha$ (all at $z=0.165$) only, and in a second fit, invoking
  a more complex mass model based on the previous conclusions, we also
  include SWc ($z=0.75$) and A901b/2-3 ($z=0.8,0.75$) as possible
  carrier of mass. The fit results can be found in Table
  \ref{tab:nfwfit}. Comparing the model fits with and without 3D
  structure in projection shows a slight drop in the mass estimates
  for A901b and SWa/b but this decrease is not statistically
  distinguishable from the simple mass model.  This demonstrates that
  previous 2D mass estimates in the literature may have overestimated
  mass measures but not significantly within the statistical errors.
  This result is expected based on Fig. \ref{fig:los} where single
  components are as equally well fit as a multiple component model
  along these lines of sight.}

{There is also the possibility that the observed features in the
  background are actually belonging to the foreground supercluster and
  are moved to higher redshift in the Wiener map owing to the noise in
  the lensing signal; their mass is relatively small (Table
  \ref{tab:nfwfit}) in the light of Fig. \ref{fig:zresolution}. In
  this case, the match with stellar mass at higher redshift would be a
  coincidence.}

We have presented a method that can be used to account for the radial
smearing of the 3D mass map and produce full redshift probability
distributions along individual l.o.s.. In future work an alternative,
potentially more fruitful method could be to first identify
significant lensing detections in a 2D map and then produce smaller
scale 3D reconstructions in these interesting regions. The benefit of
this method would be that each 3D reconstruction could be optimised
for each source by tuning the $\alpha$ parameter in the Wiener
reconstruction (see Sect.~\ref{sec:alpha}) depending of the mass and
redshift of the source. In addition, information from measurements of
higher order weak lensing flexion distortions could be included in the
analysis to boost the signal to noise in the reconstruction
\citep{Velander}. Moreover, for deeper surveys, which have reliable
photometric redshifts down to $z\sim2$ or more, we also anticipate an
moderately improved mass density recovery at higher redshift based on
simulations (Fig. 13 in STH09).

Throughout this paper, we have neglected errors caused by the
intrinsic alignment of galaxies, which for this analysis would be
dominated by the 'GI' term, the anti-correlation of sheared background
sources with the intrinsic shape of foreground galaxies.  Based on
analysis of simulations and data, the systematic error contributed by
this astrophysical effect is expected to be low on average
\citep{2010A&A...523A...1J} and probably negligible for the results
presented in this paper.  That said, the intrinsic alignment of
galaxies in cluster environments has yet to be studied in data but has
been explored using analytical prescriptions
\citep{2010MNRAS.402.2127S}.  If cluster galaxies strongly align as
they infall this could result in a strong GI term that could effect
our results by lowering the lensing signal measured behind the
foreground cluster.  The opposite however could also occur whereby
galaxy merging events in the dense cluster environment randomise the
intrinsic ellipticities resulting in a very weak GI term.  As
discussed in STH09, the contribution of GI and II correlations to the
shear field can in principle be incorporated into the minimum variance
estimator for the 3D map. At this stage, however, we expect this
effect to be small. This uncertainty is something that will be
investigated further using the STAGES data in future work.

{The driving motivation behind the development of our 3D
  reconstruction technique was to enable an unbiased 3D comparison of
  mass and light.  Dark haloes for example would only be detected in
  this manner.  However with the detailed analysis of noise and the
  radial p.s.f in the 3D lensing reconstructions presented in this
  paper, we have demonstrated how inherently noisy the process is.
  Even with one of the best HST data sets and a very massive
  foreground system we have found it difficult to reconstruct the 3D
  mass distribution at any high degree of spatial and redshift
  resolution using weak lensing alone, especially at higher redshift.
  Given the limitations of the method to resolve only the most massive
  structures in 3D the future direction for the application of this
  method should be to reconstruct larger scale structures in the 3D
  density field using more heavily spatially smoothed data.  With
  upcoming wide area surveys such as CFHTLenS, DES and VST-KIDS, which
  are several orders of magnitude wider than the STAGES survey
  presented in this paper, we can expect higher quality 3D resolution
  reconstructions on degree scales because on these scales the
  significance of modes in a 3D mass density reconstruction are
  increased (STH09).}

\section*{Acknowledgements}

We thank Peter Schneider and Jan Hartlap for useful discussions.  We
also thank the extreme snowy weather conditions that kept PS and CH
snowed in at the Royal Observatory, forcing them to finally finish
this paper.  PS acknowledges supported by the European DUEL
Research-Training Network (MRTN-CT-2006-036133) and by the Deutsche
Forschungsgemeinschaft under the project SCHN 342/7--1. CH
acknowledges support from the European Research Council under the
European Communities Seventh Framework Programme, ERC grant agreement
number 204185. TS acknowledges support from the Netherlands
Organisation for Scientific Research (NWO). MG was supported by an
STFC Advanced Fellowship and also acknowledges funding by the STAGES
NASA grant. KJ is supported by the Emmy Noether-Programme of the
German science foundation DFG. BH is grateful for support from the
Science and Technology Facilities Council (STFC).  Support for STAGES
was provided by NASA through GO-10395 from STScI operated by AURA
under NAS5-26555.

\appendix

\section{The STAGES lensing analysis}

This Appendix documents the steps taken to confirm the reliability of
the weak lensing shear catalogue that forms the basis of the 3D
reconstruction presented in Sect.~\ref{sec:results}.

\subsection{Shear catalogue: Sample I and Sample II}
\label{sect:shearcats}

The HST STAGES data were analysed independently by two authors (CH and
TS).  This involved fully independent data reduction of the Advanced
Camera for Surveys (ACS) observations, independent source extraction,
p.s.f. modelling and independent shear measurement pipelines based on
the \citet{KSB}(KSB) method.  The full HST lensing analysis is
detailed in \citet{Heymans08} (for CH) and
\citet{Schrabback07,schrabback2010} (for TS) with both KSB-based shear
measurement pipelines compared in Table A1 of \citet{STEP1}.

The main differences between the two lensing analyses are the p.s.f.
modelling method and the galaxy selection.  CH assumes the p.s.f.
temporal variation is slow over time and splits the HST data in 7
groups to achieve good temporal sampling of the p.s.f.. The resulting
semi-time dependent p.s.f. model is then applied to the data.  CH also
chooses to maximise the signal-to-noise in the lensing catalogue
fitting the isotropic correction term $P^\gamma$ as a function of
galaxy size (see \citet{Heymans08} for more details).

In the analysis of the dense stellar fields, TS has shown the HST ACS
p.s.f. can vary rapidly with time.  To account for this, TS employs a
fully time-dependent correction for the spatially and temporally
varying p.s.f. based on individual exposures, using a p.s.f. template
library obtained from stellar fields \citep{Schrabback07}.  In
addition, TS applies corrections for charge transfer inefficiency and
signal-to-noise dependent shear calibration bias as detailed in
\citet{schrabback2010}.  In this analysis of the STAGES data TS
employs conservative galaxy selection criteria including a high cut on
the shape measurement signal-to-noise ratio \citep{Erben2001}
\mbox{$\mathrm{S/N}>2.7$}.

The resulting catalogues differ in the number density of objects and
the noise on the shear measurement quantified through the root-mean
square variation of the measured galaxy ellipticities. CH, hereafter
Sample I, contains 65 galaxies per square arcmin with $\sigma_\epsilon
= 0.26$.  TS, hereafter Sample II contains 34 galaxies per square
arcmin with $\sigma_\epsilon = 0.29$.

\subsection{Photometric redshifts: COMBO-17 sample}

The STAGES HST data is complemented by the 17-band optical COMBO-17
survey.  Using a combination of broad and narrow-band filters,
COMBO-17 uses template-fitting to estimate photometric redshifts
\citep{Wolf04}.  For the A901/902 field the redshift error in units of
$\sigma_z/(1+z)$ is estimated at $\approx 0.02$ to $R<23$ with an
increase to $0.035$ in the interval $R=[23,24]$
\citep[see][]{Gray09}. Outlier rates also increase with magnitude: we
estimate roughly $1\%$ outliers with $|\Delta z|/(1+z)>0.10$ at $R<22$
and $7\%$ outliers with $|\Delta z|/(1+z)>0.15$ at $R=[23,24]$. Above
a magnitude $R=24$ we do not use COMBO-17 redshift estimates due to
rapidly increasing errors and outlier rates.

\begin{figure}
 \begin{center}
   \psfig{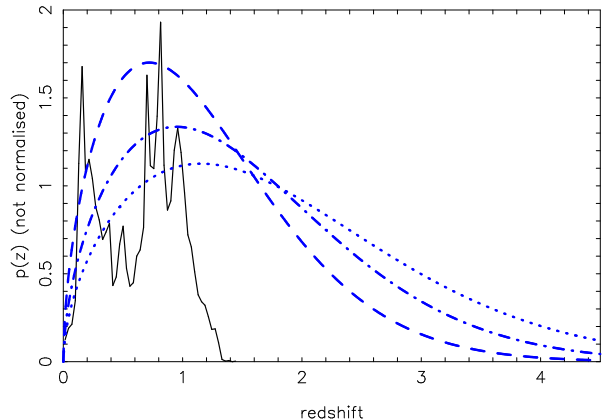}
 \end{center}
 \caption{\label{fig:redshifthist} Distribution of sources in
   subsamples as used in the 3D reconstruction. The solid line
   shows the distribution of photometric redshifts for sources with
   photo-$z$'s, the other lines are estimated distributions for three
   magnitude bins (Sample I; Sample II is only marginally different)
   including all other sources: $23\le m_{\rm F606W}<25$ (dashed),
   $25\le m_{\rm F606W}<26$ (dashed-dotted) and $26\le m_{\rm
   F606W}<27.5$ (dotted). Sources with photometric redshifts are
   further sub-divided into $21$ photo-z bins. Distributions have
   arbitrary normalisations.}
\end{figure}

Owing to the width of the lensing kernel in redshift space, the
accuracy of the \mbox{COMBO-17} redshifts are more than sufficient.
Redshift estimates, however, only exist for a small fraction of the
galaxies ($13$ galaxies per square arcmin for Sample I and $10$
galaxies per square arcmin for Sample II).  For the remaining galaxies
we assign a magnitude dependent redshift distribution given by
\citet{Schrabback07} (see Fig. \ref{fig:redshifthist}). All this
information is combined for the final mass map.

The stellar mass estimates used in this analysis derive from low
resolution 17-band spectra fits to parameterised star formation
history models as described in \citet{2006A&A...453..869B}.

\subsection{Quality control; comparison of E and B mode signals to
 pure noise maps}
\label{sect:noisepeaks}

\begin{figure}
 \begin{center}
   \psfig{file=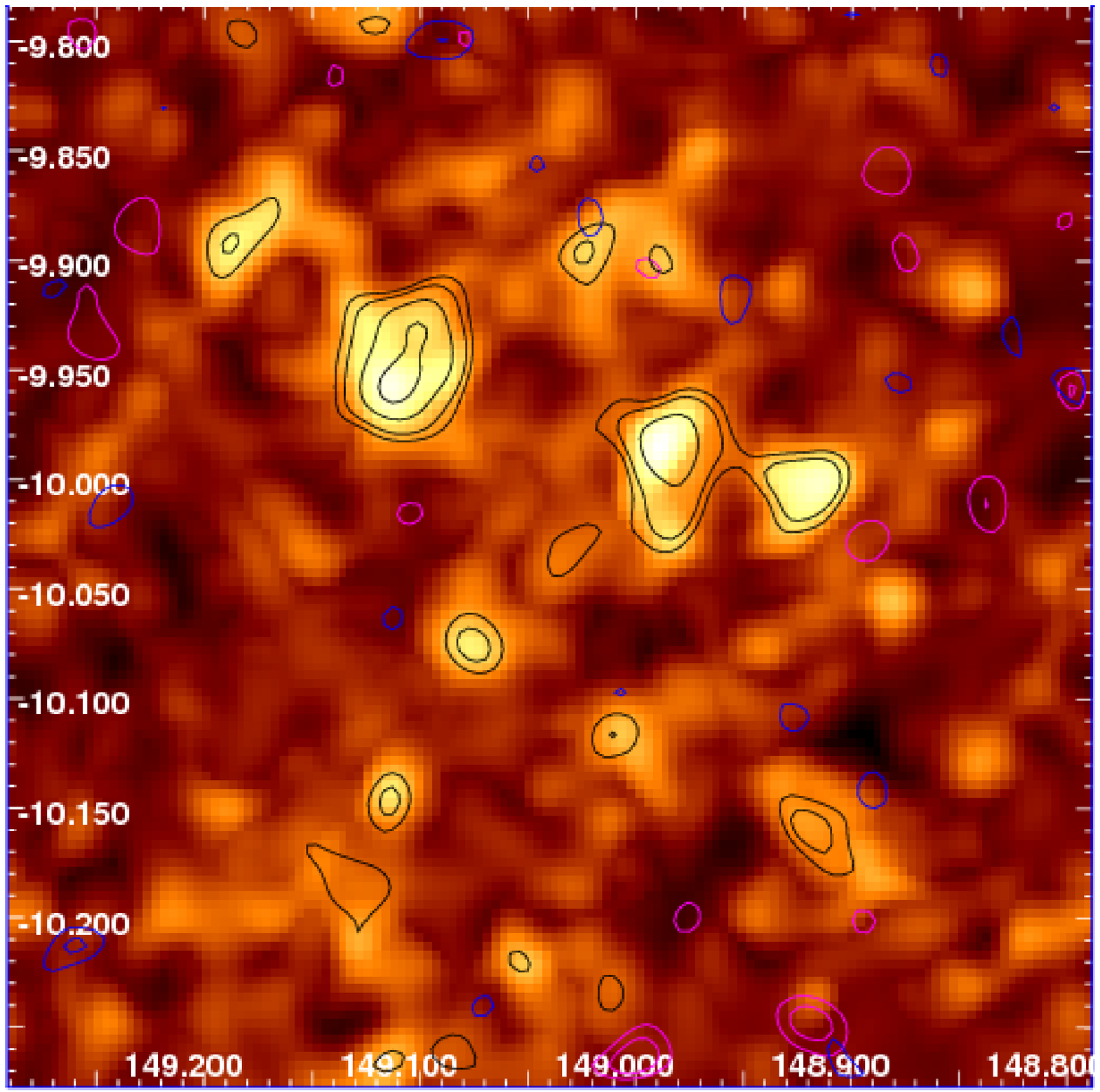,width=87mm,angle=0}
 \end{center}
 \caption{\label{fig:stages3dtim} Comparison of the mass reconstruction from
   SAMPLE I and SAMPLE II.
   On-sky projection of Wiener map
   using SAMPLE II (intensities in the background are maximum density
   contrast along l.o.s.). Black solid lines are signal-to-noise
   levels $2.5\sigma$, $3\sigma$, $4\sigma$, $5\sigma$ and $6\sigma$
   of the E-mode map. The blue contours show the signal-to-noise in
   the B-mode map. For comparison, the magenta contours encircle
   B-mode signal-to-noise peaks in the SAMPLE I map. The values on the
   $x$-axis denote the right ascension in degrees and the declination
   for the $y$-axis (J2000.0).}
\end{figure}

{A reconstruction of signal and noise, based on the shear Sample
  I and shear Sample II, is carried out following STH09 with the
  resulting map is shown in Fig. \ref{fig:stages3dmap} for Sample I
  and Fig. \ref{fig:stages3dtim} for Sample II.}

We find that both samples detect the same significant structures at
the positions of the known galaxy clusters A901a, A901b and A902, in
addition to a detection of a group of galaxies called the South West
group (SW) as seen before in earlier 2D reconstructions of the
projected matter density \citep{Heymans08,Gray02}.  As expected from
the galaxy number counts in the two samples, the signal-to-noise in
the maps is higher for Sample I than Sample II.  All structures that
differ between the maps are at a level consistent with noise.

For an analysis of the systematics, we count the number of peaks and
troughs in the E- and B-mode S/N 3D maps given signal-to-noise range
and compare this to the expected numbers in pure noise maps. The noise
statistics are obtained from the randomised data. In order to have an
easily computable and objective statistic for ``peaks'' and
``troughs'', we count the number of grid pixels in the Wiener map
under the condition that the pixels are local maxima within a cuboid
volume centred on the pixel. The size of the cuboid is $\pm2$ pixels
in transverse and $\pm2$ lens-planes in radial direction. For
S/N-thresholds $\gtrsim3\sigma$ this definition fits reasonably well
the number of peaks one would select by eye. Troughs are considered
negative peaks (local minima with negative signal-to-noise).

\begin{figure}
  \begin{center}
    \psfig{file=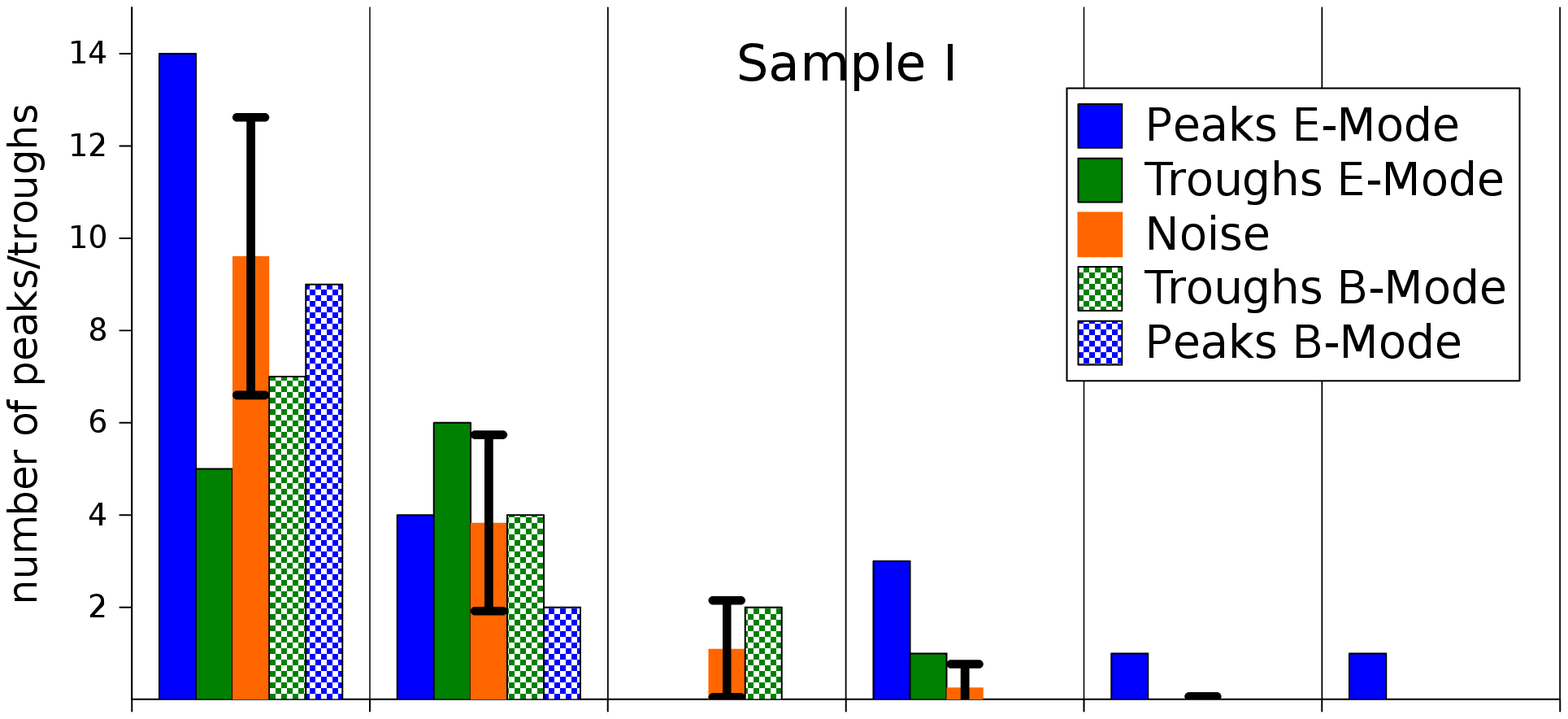,width=80mm,angle=0}\\
    \psfig{file=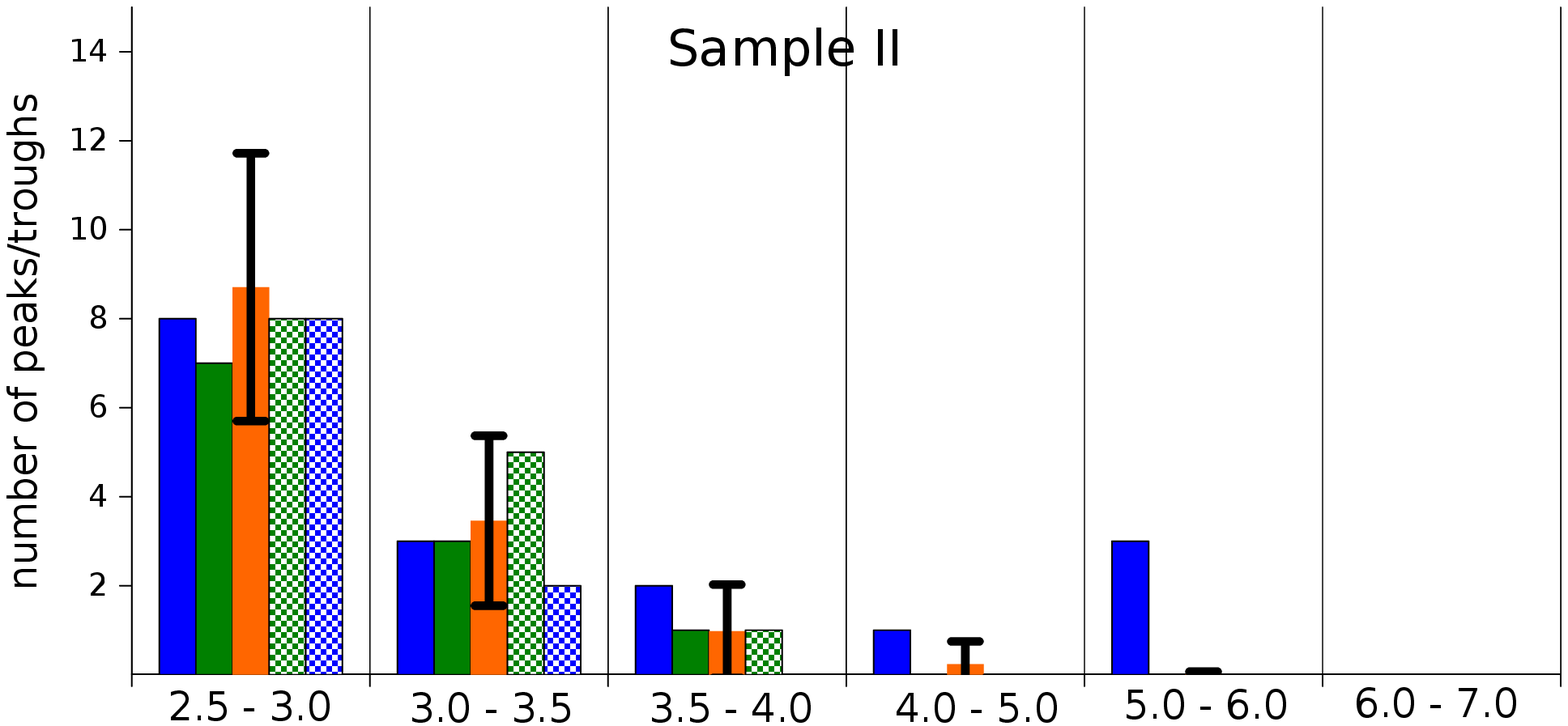,width=80mm,angle=0}\\
  \end{center}
  \caption{\label{fig:peakstat} Number of peaks and troughs within a
    given signal-to-noise bin (\emph{top panel}: Sample I,
    \emph{bottom panel}: Sample II). The orange bar in the middle
    shows the mean number of noise peaks and their r.m.s. variance as
    determined from noise realisations ($10^3$), solid bars are
    statistics of the E-mode map, whereas shadowed bars show the
    statistics of the B-mode map. The exact definitions of pixel peaks
    and troughs is given in the text. {Note that the noise
      statistics of peaks and troughs is within numerical errors
      identical so that both are represented by the same orange bar.}}
\end{figure}

Fig. \ref{fig:peakstat} shows the results for Sample I and Sample
II. Clearly, for thresholds $\gtrsim3.5\sigma$, E-mode peaks (troughs)
are above the average noise levels and above $1\sigma$-noise variance
for $\gtrsim4.0 \sigma$, whereas B-mode counts are consistent (Sample
I) with the noise expectation. From the noise realisations, we
estimate the covariance of the noise counts which is used for a
$\chi^2$-test, peak and trough counts combined, to check consistency
with noise. For Sample I we find $\chi^2/{\rm d.o.f.}=177.4$ (E-mode
map) and $\chi^2/{\rm d.o.f.}=0.40$ (B-mode map), showing that the
E-mode map is inconsistent with noise, while the B-mode map is
consistent with noise. Corresponding values for Sample II are
$\chi^2/{\rm d.o.f.}=220.0$ (E-mode) and $\chi^2/{\rm d.o.f.}=0.29$
(B-mode) leading us to a similar conclusion.

We therefore find the B-modes for both Samples to be low and
consistent with noise.This is potentially a surprise for Sample I,
bearing in mind the rapid variations in p.s.f. ellipticity seen in
\cite{Schrabback07} but unaccounted for in the Sample I analysis.  The
success of the semi-time dependent p.s.f. subtraction method of Sample
I may be explained by the fact that HST was significantly under-focus
during the STAGES observations with the result of a relatively stable
p.s.f., once averaged over intra-orbit variations. Owing to the larger
Sample I, there are peaks up to $7\sigma$, unseen for Sample II.

Note that in the high-S/N regime the number of peaks or troughs
becomes rare and statistically Poissonian (one or no peak) which may
not be properly reflected by the symmetric error bars in the figure.

\subsection{Redshift Scaling}

\begin{figure}
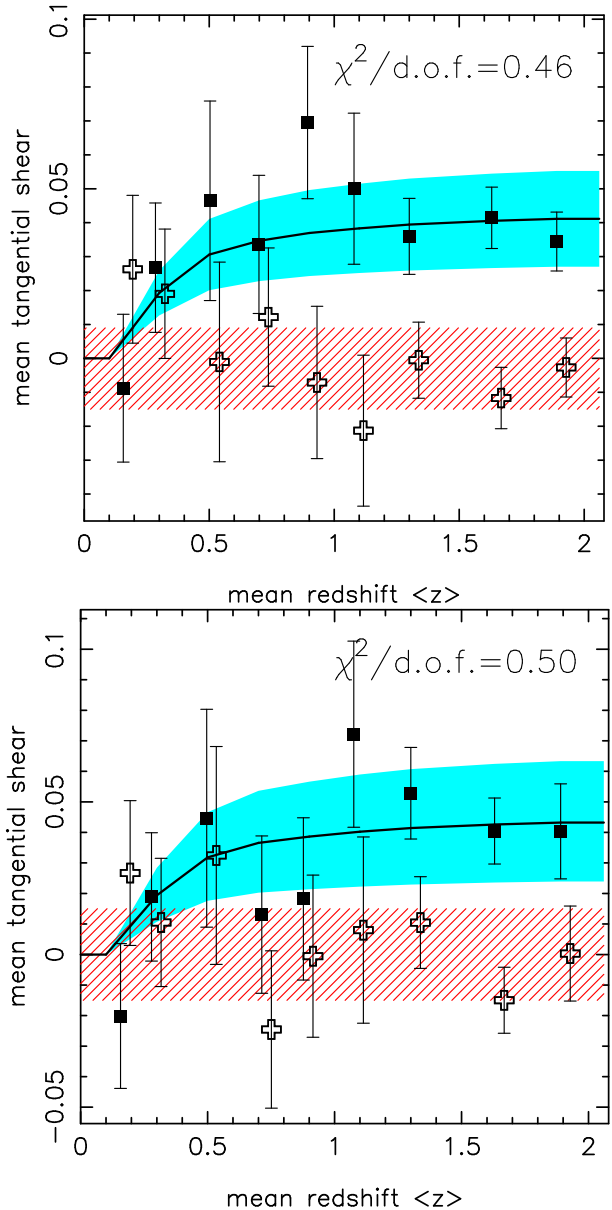

 \begin{center}
   \psfig{file=fig15.ps,width=80mm,angle=-90}\\
   \psfig{file=fig16.ps,width=80mm,angle=-90}
 \end{center}
 \caption{\label{fig:zscaling} {Test of redshift-scaling of the
     lensing
     signal in the data.} Mean tangential shear as function of source
   redshift within an aperture of radius $2$ arcmin (A901a and A901b)
   or $1.4$ arcmin (A901$\alpha$, A902/1 and SWa/b) centred on the
   peak centres; the signal for all four apertures is averaged. The
   sources are binned in redshift for the samples with photometric
   galaxy redshifts ($z\le1.2$). The three highest $z$-bins belong to
   the magnitude samples for which redshifts correspond to the mean as
   determined from the redshift distribution of sources. With most
   matter inside the aperture located at the same redshift, the mean
   tangential shear (solid points) should scale according to the black
   solid line in the plot. The shaded region around the line is the
   $1\sigma$-confidence region of the amplitude. Open crosses denote
   the average radial shear inside the aperture, which should be
   vanishing. The shaded red area marks the $1\sigma$-area for the
   mean B-mode over all redshifts. \emph{Top panel:} Sample I,
   \emph{bottom panel:} Sample II.}
\end{figure}

Recent lensing analysis of ground-based data using the KSB method
found errors in the expected redshift scaling of the measured cosmic
shear signal \citep{Kilbinger09}.  The reason for this bias is unknown
but is thought to lie in a weak signal-to-noise and galaxy-to-psf size
dependent calibration bias inherent to shear measurement methods
\citep{STEP2,GREAT08,Semboloni09}.

As a further quality check we test the redshift scaling of the
KSB-measured space-based lensing signal for Sample I and II.  We
select a set of peaks that are known to be at redshift $z\approx0.165$
(A901a/b/$\alpha$, A902/1 and SW/a/b).  For each peak the mean
tangential shear within an aperture, of radius between one and two
arcmin, is measured from the data:
\begin{equation}
 \bar{\gamma}_{\rm t}(\vec{\theta}_{\rm c},\bar{z})=
 -\sum_i\frac{\vec{\theta}_i^\ast-\vec{\theta}_{\rm c}^\ast}
 {\vec{\theta}_i-\vec{\theta}_{\rm c}}\epsilon_i\;,
\end{equation}
where $\vec{\theta}_{\rm c}$ is the l.o.s. direction of the aperture
centre, $\vec{\theta_i}$ the angular position of a source in the
common complex notation and $\epsilon_i$ the complex ellipticity of
the source. Only sources within the aperture and within some redshift
range are considered. The statistical uncertainty of the estimator for
the mean shear is the variance of the mean as obtained from the
sources included in the sum. The mean redshift, $\bar{z}$, of included
sources determines the redshift associated with $\bar{\gamma}_{\rm
  t}$. The statistics from all lensing peaks is combined and weighed
by their statistical uncertainties. Thus we are stacking the cluster
signals to increase the signal-to-noise of this test.

Fig. \ref{fig:zscaling} shows aperture statistics for Sample I and
Sample II by plotting the real parts of $\bar{\gamma}_{\rm t}$
(E-mode) and the imaginary parts (B-mode; radial shear) separately. If
the lensing signal within the apertures stems in essence only from
structure located at the cluster redshift $z\sim0.165$, we expect a
redshift scaling dictated by the lensing efficiency (with fixed lens
distance): \begin{equation}
 \label{eq:zscaling}
 \bar{\gamma}_{\rm t}(\bar{z})\propto 
 \Ave{\frac{f_{\rm k}(w_{\rm
 s}-w_{\rm h})} {f_{\rm k}(w_{\rm s})}}_{w_{\rm s}}\approx
\frac{f_{\rm k}(\ave{w_{\rm
 s}}-w_{\rm h})} {f_{\rm k}(\ave{w_{\rm s}})}\;,
\end{equation}
where $f_{\rm k}(w)$ is the angular diameter distance as function of
radial comoving distance $w$; $w_{\rm h}$, here constant, and
$\bar{w}_{\rm s}$ are the halo and average source distance,
respectively. The last step is only approximately correct. This
approximation is used for the fit in Fig. \ref{fig:zscaling}.

This behaviour is independent of the intrinsic density profiles of the
clusters, albeit its amplitude depends on the cluster details. As we
are solely interested in verifying the correct redshift scaling
behaviour, we fit Eq. \ref{eq:zscaling} with the amplitude as unknown
variable to the E-mode data points which {for} $\chi^2/{\rm
  d.o.f.}=0.46(0.50)$ is an excellent fit to Sample I (Sample II).

B-modes should be consistent with zero for all redshifts. This we test
by fitting a straight horizontal line, with unknown amplitude, to the
average radial shear. We find that the fit is both consistent with
zero on a $1\sigma$-level and well fitted by a model with
redshift-independent signal, yielding $\chi^2/{\rm
  d.o.f.}=0.97(0.87)$.

\subsection{Conclusion}

In conclusion, we find both shear catalogues to have a low level of
systematics in the B-mode channel consistent with noise and a
redshift-scaling of the shear signal that is consistent with
$\Lambda$CDM cosmology.  Moreover, we find no significant differences
between the E-mode channel maps of Sample I and Sample II. As Sample I
yields an overall more significant map owing to the higher number
density of sources, we present results in the main section of the
paper only from the data of Sample I (CH).

\bibliographystyle{mn2e}
\bibliography{STAGES3D}

\bsp
\label{lastpage}

\end{document}